%% file: Main.tex
\documentclass[a4paper,11pt]{article}
\pdfoutput=1 

\usepackage{jcappub} 

\usepackage[T1]{fontenc} 
\usepackage[hang, raggedright, nooneline]{subfigure}
\usepackage{booktabs, diagbox}

\newcommand{\Equ}[1]{Equ.\,(\ref{#1})}
\newcommand{\Equation}[1]{Equation\,(\ref{#1})}
\newcommand{\Fig}[1]{Fig.\,\ref{#1}}
\newcommand{\Figure}[1]{Figure\,\ref{#1}}
\newcommand{\Tab}[1]{Table\,\ref{#1}}
\newcommand{\Table}[1]{Table\,\ref{#1}}
\newcommand{\Sec}[1]{Section\,\ref{#1}}

\title{\boldmath CRPropa 3.1 -- A low energy extension based on stochastic differential equations}

\author[a, 1]{Lukas Merten}
\author[a, 1]{Julia Becker Tjus}
\author[b, 1]{Horst Fichtner}
\author[a, 1]{Björn Eichmann}
\author[c]{G\"unter Sigl}

\affiliation[a]{Ruhr-Universität Bochum, Theoretische Physik IV: Plasma-Astroteilchenphysik,\\Universit\"atsstrasse 150, 44801 
Bochum, Germany}
\affiliation[b]{Ruhr-Universität Bochum, Theoretische Physik IV: Weltraum- und Astrophysik, \\Universit\"atsstrasse 150, 44801 
Bochum, Germany}
\affiliation[c]{University of Hamburg, II Institut für Theoretische Physik\\ Luruper Chaussee 149, 
22761 Hamburg, Germany}
\affiliation[1]{Ruhr Astroparticle- and Plasmaphysics Center (RAPP Center)}

\emailAdd{lukas.merten@rub.de}
\emailAdd{julia.tjus@rub.de}
\emailAdd{hf@tp4.rub.de}
\emailAdd{eiche@tp4.rub.de}
\emailAdd{guenter.sigl@desy.de}

\keywords{Cosmic Rays, Transport, Stochastic Differential Equation, Anisotropic Diffusion}

\abstract{
The propagation of charged cosmic rays through the Galactic environment influences all aspects of 
the observation at Earth. Energy spectrum, composition and arrival directions are changed due to deflections 
in magnetic fields and interactions with the interstellar medium. Today the transport is simulated 
with different simulation methods either based on the solution of a transport equation 
(multi-particle picture) or a solution of an equation of motion (single-particle picture).

We developed a new module for the publicly available propagation software CRPropa 3.1, where we 
implemented an algorithm to solve the transport equation using stochastic differential equations. 
This technique allows us to use a diffusion tensor which is anisotropic with respect to an arbitrary 
magnetic background field. The source code of CRPropa is written in C++ with python steering via SWIG which makes it easy to use and computationally fast.

In this paper, we present the new low-energy propagation code together with validation procedures that are developed to proof the accuracy of the 
new implementation. Furthermore, we show first examples of the cosmic ray density evolution, which 
depends strongly on the ratio of the parallel $\kappa_\parallel$ and perpendicular $\kappa_\perp$ 
diffusion coefficients. This dependency is systematically examined as well the influence of the particle rigidity on the diffusion process.
}

\begin{document}
\maketitle
\flushbottom

\include{Introduction}

\include{SDE}

\include{Validation}

\include{Example}

\include{Summary}

\appendix
\include{Appendix}

\acknowledgments

The authors would like to thank the whole CRPropa development team, especially G.\ Müller, A.\ Dundovic and D.\ Walz, for their persistent help with the implementation of 
the algorithm. We also appreciate discussions with A.\ Kopp and M.S.\ Potgieter. Some of this work was made possible by the following python packages: NumPy and SciPy 
\cite{van_der_Walt_2011}, Pandas \cite{McKinney_2012}, matplotlib \cite{Hunter_2007} and IPython \cite{Perez_2007}. We would further like to acknowledge support from the RAPP 
Center (Ruhr Astroparticle and Plasmaphysics Center sponsored by the MERCUR project St-2014-0040) and from the Research Department of Plasmas with Complex Interactions (Bochum). 
GS is supported by the DFG through collaborative research centre SFB 676,
by the Helmholtz Alliance for Astroparticle Physics (HAP) funded by the Initiative and
Networking Fund of the Helmholtz Association and by the Bundesministerium für Bildung
und Forschung (BMBF) through grant 05A14GU1.



%
%
%
%

\bibliography{literature}
\bibliographystyle{JHEP}

\end{document}

%% file: Introduction.tex
\section{Introduction}
\label{sec:intro}

While the field of astroparticle physics has advanced significantly during the past decade coming closer to an identification of the sources, the origin of both Galactic and 
extragalactic cosmic rays is still not fully resolved. Modern 
experiments like IceCube \cite{icecube_concept, icecube2014}, Imaging Air Cherenkov Telescopes (H.E.S.S. \cite{hess_telescopes2003}, MAGIC \cite{magic2004}, VERITAS 
\cite{veritas2004}), the  Pierre Auger Observatory (PAO) \cite{Abraham2010}, the Telescope Array (TA) \cite{ta_spectrum2015, ta_composition2015} or AMS-02 \cite{Weng2016} 
challenge theories and 
simulations with a 
high accuracy of the observed data. Especially the transition between the Galactic and 
extragalactic part of the energy spectrum is not fully understood. There are different simulation frameworks available for low-energy, Galactic and high-energy extragalactic 
propagation that meet a high standard already today \cite{GALPROP, Evoli2008, Kissmann2014}. One unified simulation framework, however, that can 
broadly and consistently describe all three major characteristics of the observations --- namely 
chemical 
composition, energy spectrum and anisotropy level of the arrival directions --- is not yet developed. 

In particular, there are two different fundamental approaches for the modeling of cosmic ray (CR) transport: (1) The solution of the transport equation enables a 
approximate description in different environments, and thus has the major disadvantage that it is necessary to model the diffusion of the particles via a diffusion tensor, 
often reduced to a one dimensional diffusion coefficient. Here, also a variety of (semi-) analytic theories exist. These are e.g.\ the leaky box or more complex models (see e.g. 
\cite{V.S.1990, eichmann2016, yoast_hull2013, schlickeiser_jenko2010}). These models are able to describe the global shape of the spectrum but cannot fully reproduce the latest 
precision measurements. (2) The solution of the equation of motion, which is a very precise, but also a very CPU-time-consuming ansatz, only being usable at the highest energies, 
where it is feasible with respect to the computation times.

\paragraph{Transport Equation} On Galactic scales, the solution of a set of transport equations is commonly used, describing the flux for the different particle species and the 
change of their energy spectrum through processes like diffusion in space and momentum, advection and adiabatic cooling. A typical representation of the transport equation is 
given here:
\begin{align}
  \frac{\partial n}{\partial t} +  \vec{u} \cdot \nabla n = \nabla\cdot(\hat{\kappa}\nabla n) + 
\frac{1}{p^2}\frac{\partial}{\partial p}\left(p^2\kappa_{pp}\frac{\partial n}{\partial p}\right) 
+ \frac{1}{3}\left(\nabla \cdot \vec{u}\right)\frac{\partial n}{\partial \ln p} + S(\vec{x}, p, t) 
\quad . \label{eq:DiffusionEquation}
\end{align}
This is the so called Parker transport equation (augmented with a term describing momentum diffusion) which is a simplified version of Fokker-Planck transport equation.
It is believed to be a good description of the particle transport problem in diffusive regimes 
(e.g.\ \cite{GALPROP}). Here, $n$ is the particle density, $\vec{u}$ is the advection speed, 
$\hat{\kappa}$ is the
spatial diffusion tensor, $p$ is absolute momentum, $\kappa_{pp}$ is the momentum diffusion 
coefficient used to describe re-acceleration and $S(\vec{x}, p, t)$ describes the sources of 
cosmic rays. For the case of an entire nuclear network, as it is present in the cosmos, \Equ{eq:DiffusionEquation} can be extended to include catastrophic losses and gains from 
spallation (interaction with background matter) and nuclear decay. Furthermore, formulations of continuous losses from interactions with the magnetic field (e.g.\ synchrotron 
radiation) can be developed also for single species simulations.


Nowadays, there exist three major simulation tools for the Galactic cosmic ray transport: GALPROP 
\cite{GALPROP}, DRAGON \cite{Evoli2008} and PICARD \cite{Kissmann2014}. GALPROP is probably the 
most widely used program but does not 
provide a method to use anisotropic, space dependent diffusion tensors, yet. Nonetheless, it is an 
excellent instrument to describe spallation and a variety of loss processes. DRAGON, which is based 
on 
the same principles as GALPROP, addresses the anisotropic diffusion problem among other things. It is known that 
the diffusive behavior along and perpendicular to the mean magnetic field line differ from each 
other (e{.}g{.} \cite{Shalchi2009}). A more sophisticated numerical 
approach to solve the transport equation is PICARD. The biggest difference is that it uses 
multi-grids and 
provides a solver explicitly designed to yield the stationary solution of the transport equation. 

Another technically different approach to solve the transport equation, commonly used for the 
propagation 
of cosmic rays inside the 
heliosphere \cite{Busching2011, Effenberger2012}, makes use of the equivalence of parabolic partial 
differential equations and a
corresponding set of stochastic differential equations, known from stochastic integrals and Ito 
calculus (see \Sec{sec:SDE} and \cite{gardiner2009}). First attempts to 
apply this method also to the Galactic propagation of cosmic rays are made in \cite{Muraishi2009, 
Effenberger2012a, Miyake2015, Kopp2014}. This concept is part of the methods 
introduced for Galactic 
propagation in this paper and will be discussed in detail in \Sec{sec:SDE}. 

\paragraph{Equation of Motion} In CRPropa \cite{kampert2013, Batista2016} a single-particle approach is used for description of the transport of extra-galactic CRs.  CRPropa\,3.0 
propagates single particles forward in time through the numerical integration of the equation of motion 
\cite{Batista2016}. This opens up many new possibilities for the simulation. 
For example, no assumptions on the diffusion tensor have to be made, but arbitrary magnetic field configurations can be used. This makes the simulation more 
fundamental. In addition things like tracing of distinct particle paths and a stochastic treatment 
of interactions is possible. However, this method is not applicable to Galactic transport problems, 
because it is computationally too time consuming for simulations using particles with energies below 
a few $10\,{\rm PeV}$. 

The energy range between the knee at ~$10^{15}\,{\rm eV}$ \cite{Wiebel-Sooth1997} and the ankle at ~$10^{18.5}\,{\rm eV}$ \cite{Abraham2010} in the cosmic ray spectrum is called 
transition region because there the source distribution shifts probably from Galactic to extragalactic sources (see e.g.\ \cite{allard2007} for a recent review). It is 
unclear where the influence of the extragalactic part of the spectrum begins to dominate. 
Furthermore, it would be interesting to decompose these two compositions in the knee-to-ankle 
region to learn more about possible sources with energies up to EeV energies.
Up to now no open-access program exists which models the transition region as a combination of Galactic and 
extragalactic influences. Individual models have been presented describing a possible contribution of Galactic  and extragalactic  cosmic rays in this transition region 
\cite{stanev1993, bbr2009, hoerandel2004, giacinti2014, Thoudam2016}. 
In this paper, we introduce an extension of the public CRPropa code that enables the user to perform cosmic ray simulations at hundreds of PeV to ZeV energies with the 
single-particle propagation method, but also the modeling of TeV-PeV energies with the low energy add-on that we present in this paper using stochastic differential equations for 
the multi-particle propagation. This code is 
open-access\footnote{\texttt{https://github.com/CRPropa/CRPropa3/releases}} and can be used by the entire community \cite{Batista2016}.

\subsection{The Structure of CRPropa}
\label{ssec:ProgramStructure}

In this section we explain the program structure of CRPropa 3.1 and how the new \texttt{DiffusionSDE} module is related to the other modules. CRPropa 3.1 is build of several 
independent 
modules which can be used to create a user defined simulation setup. This structure makes it easy to extend CRPropa with new modules, e.g.\ with new interaction modules.

The central data structure in CRPropa is the so called candidate. This object holds all necessary information about the (pseudo-) particle such as energy, direction, particle 
type, position and many more attributes. These attributes are altered and processed by the 
\texttt{ModuleList} during the propagation procedure. The \texttt{ModuleList} is basically a list 
containing all simulation modules chosen for a specific simulation. The user can load different kinds 
of modules into the \texttt{ModuleList}:
\begin{enumerate}
 \item \emph{Boundary modules}: Deactivate (stop simulation) the candidate if the boundary condition is met.
 \item \emph{Source modules}: Create new candidates and inject them into the simulation.
 \item \emph{Interaction modules}: Take interactions into account, e.g.\ proton-photon-interaction, radioactive 
decay, etc.
 \item \emph{Deflection modules}: Change the position of the candidate according to the equation of motion (\texttt{PropgationCK}) or transport equation (\texttt{DiffusionSDE}).
 \item \emph{Observer modules}: Detect particle if a certain condition is fulfilled.
 \item \emph{Output modules}: Save the simulation results to a text- or hdf-file.
\end{enumerate}

\begin{figure}[htb]
\centering
 \includegraphics[height = .35\textheight]{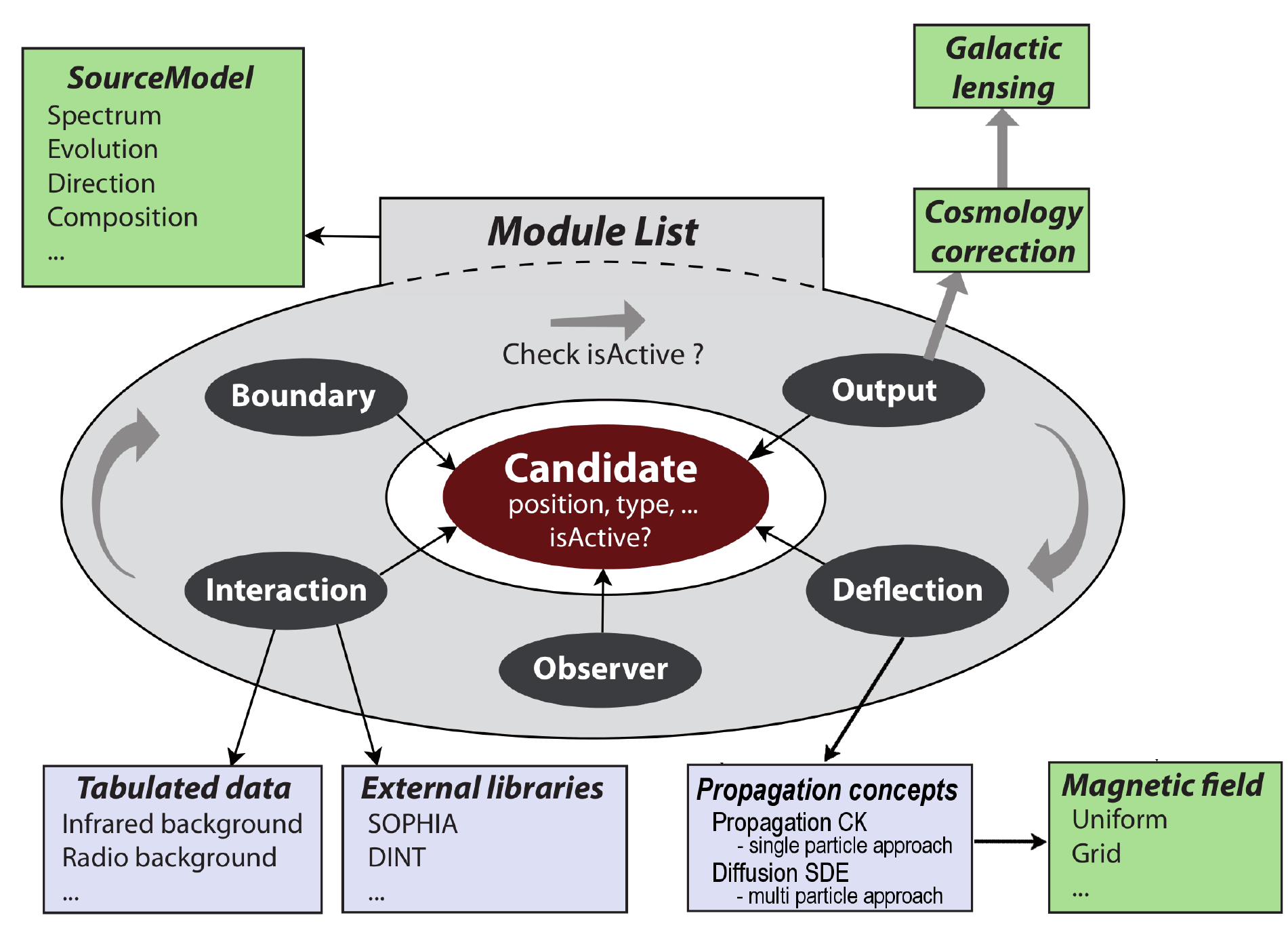}
 \caption{The modular structure of CRPropa allows for an easy extension of the program. In every 
propagation cycle the \texttt{isActive} flag of the candidate is checked and serve as the break 
condition. The \texttt{DiffusionSDE} module can be used instead of the \texttt{PropagationCK} 
module to update the (pseudo-) particle position. Figure taken from \cite{Batista2016} and adjusted with kind permission.}
 \label{fig:CRPropaStructure}
\end{figure}

\Figure{fig:CRPropaStructure} shows the simulation process. The modules, except for the source module, are called successively and act on a candidate  with each integration time 
step $h$ as long as the candidate is active.
%

The new module \texttt{DiffusionSDE} can replace the module \texttt{PropagationCK} as needed. In particular, we recommend to use the \texttt{DiffusionSDE} module for propagation 
with an energy below 100 PeV, while \texttt{PropagationCK} should be used for samples that are dominated by energies range above 100 PeV. While these 
numbers are a good estimate for the Galaxy they may differ significantly in other surroundings.

It cannot be generally answered which module is better. The two modules simulate two fundamentally different descriptions of the propagation process. Which module suits 
best depends on the specific simulation setup. In the case of weak magnetic fields with gyro radii of the order of the coherence length of the magnetic field a description by the 
transport equation is not possible. Also in the case of vanishing turbulent magnetic fields the particle transport is not diffusive and therefore can not be described by 
\Equ{eq:DiffusionEquation}. In general, the distribution function $n$ has to be isotropic in momentum space to be described by the diffusion approximation. 

On the other hand, the propagation of particles in very turbulent magnetic fields, where the gyro radius is smaller than the coherence length of the 
field, is computing time intensive. This may prevent the generation of sufficient statistics to draw 
significant conclusions.

In addition, we want to give a short overview of the different output formats and how they can be used to create observables which can be compared to measurements. CRPropa outputs 
can 
be generated in a tracking mode, where every (pseudo-) particle position during the simulation process is recorded. This is very useful if the concrete particle trajectory is of 
interest, but obviously yields huge amounts of data if it used for a full scale simulation with billions of candidates and hundred thousands of integration time steps for each 
candidate. In 
contrast to that, most other output modes save only those candidates which are 
detected by an \texttt{ObserverModule}. Here, the cosmic ray properties (position, momentum, etc.) at the origin and at detection are stored. With the \texttt{DiffusionSDE} 
module we developed also a new \texttt{ObserverTimeEvolution} module which takes snapshots of the total cosmic ray distribution for user defined time points. 

Furthermore, not only cosmic rays but also neutral particles like $\gamma$ rays or neutrinos, which are created as secondaries during interactions, can be propagated and stored.

%% file: SDE.tex
\section[Transport model]{Transport Model and Stochastic Differential Equations}
\label{sec:SDE}

In order to extend CRPropa toward lower energies and by that to a dedicated Galactic propagation tool, we 
decided to use stochastic differential equations. This ansatz is most compatible with the single particle tracking used for the extra-galactic propagation.

Furthermore, it has some advantages over classical grid 
based methods. Since the calculation is not restricted to a grid, adaptive algorithms can be applied to 
vary the integration time step which depends on the local magnetic field geometry. This leads to a 
simulation as accurate as needed and as effective as possible. To take all possible interaction channels 
into account the grid based methods have to calculate the whole (DRAGON and GALPROP) or parts 
(PICARD) of the reaction network twice. That is not necessary following the stochastic approach, where all secondary particles are propagated as new particles.

The diffusion equation is solved in the local frame of the magnetic field line 
(see \Sec{ssec:Code}) which implies nearly no restriction on the allowed magnetic field set up. 
This means that no analytical description of the magnetic field is needed. In contrast to the 
approach in \cite{Effenberger2012a} the magnetic field direction is calculated on the fly and there 
is no need for derivatives of the vector field.

In times of huge CPU clusters the potential of parallelization is a crucial factor for a simulation 
program. The solution of the stochastic differential equation (SDE) is trivial to parallelize because the different phase space 
elements (hereafter called pseudo-particles) are independent of each other. The computation time for pure propagation of pseudo-particles scales linearly with the number of cores.

Furthermore, the simulation results can be reweigthed after the actual simulation is finished. The pseudo-particles can for example be weighted according to their energy at the source to change effectively the energy spectrum of the source distribution without the need of a new full size simulation. This makes it easy and computationally cheap to do parameter studies. It has to be mentioned that a change of the propagation parameters, such as diffusion coefficient or magnetic field configuration cannot be changed after the simulation.

\subsection{Stochastic Differential Equations}
\label{ssec:SDE}
In this section we briefly summarize the mathematical background of stochastic differential equations. Here, we only discuss the one-dimensional case which can in principle be generalized to more dimensions. Furthermore, it should be noted that the rigorous mathematical description is always just an approximation of a real physics problem, especially with respect to correlation times etc.

Stochastic differential equations have been examined since more than one hundred years ago. Starting with studies on the Brownian motion by Einstein and Langevin who derived the 
so called 
Langevin-equation independently:
\begin{align}
 \frac{{\rm d}x}{{\rm d}t} = a(x, t) + b(x, t)\xi(t) \label{eq:Langevin} \quad ,
\end{align}
where $\xi(t)$ is a ``rapidly [in time] fluctuating random term'', meaning $\xi(t)$ and $\xi(t')$ are statistically independent for $t\neq t'$ \cite{gardiner2009}. That means the random term is fully uncorrelated which can be described as:
\begin{align}
\langle\xi(t+\tau)\xi(t)\rangle&=\lim_{T\rightarrow\infty} 1/T \int_0^T \xi(t)\xi(t+\tau)\;{\mathrm d}t\\ \notag
&=\delta(\tau) \quad .
\end{align} 
Furthermore, we require the time average to vanish
($\langle\xi(t)\rangle=0$) because any drift term can be described by $a(x, t)$. \Equation{eq:Langevin} can consistently be interpreted by the 
corresponding integral equation (see e.g.\ \cite{gardiner2009}):
\begin{align}
 x(t) = x(0) + \int_{t_0}^t a[x(s), s]\;{\rm d}s + \int_{t_0}^t b[x(s), s]\;{\rm d}W(s) \label{eq:IntegralEquation} \quad .
\end{align}
Here, we used the fact that the integral of $\xi(t)$ is a Wiener process, $\int_{t_0}^t \xi(t')\,{\rm d}t'=W(t)$, which allows for the replacement of $\xi(s)\,{\rm d}s={\rm 
d}W(s)$. A Wiener Process is a Markov process and it can be interpreted as the solution of the diffusion equation with zero drift and a constant diffusion coefficient that is equal to 1. The sample paths are continuous  but not differentiable.\Equ{eq:IntegralEquation} is the prototype of a stochastic integral equation. The quantity $x(t)$ is determined by the initial condition $x(0)$ a deterministic term (first integral) and a stochastic term (second integral).  The details of the existence and uniqueness of solution of these types of equations is 
beyond the scope of this paper and we refer the reader to the detailed studies in the textbook by Gardiner \cite{gardiner2009}. We just want to emphasize that the solution is not fully determined by the starting condition --- as it is the case for ordinary differential equations --- but also depends on the particular realization of the Wiener process.  
In addition it should be mentioned that the last term of \Equ{eq:IntegralEquation} --- the stochastic integral --- can be interpreted in two ways, which differ in the choice where between the grid points $t_i$ and
$t_{i-1}$ with a time ordering $t_0\leq t_1 \leq \dots \leq t_n = t$, the stochastic part of the integral ($b[x(s), s)$] is evaluated. It can be shown that the stochastic differential equation 
is 
independent of the two, namely It\={o} and Stratonovich, 
interpretations of the stochastic integral \cite{gardiner2009}.

If a stochastic quantity $x(t)$ is described by \Equ{eq:IntegralEquation} for all $t$ and $t_0$ it obeys the It\={o} SDE given by:
\begin{align}
 {\rm d}x(t)=a[x(t), t]{\rm d}t + b[x(t), t]{\rm d}W(t) \label{eq:ItoSDE} \quad .
\end{align}
it can be concluded that the solution of the It\={o} SDE \Equ{eq:ItoSDE} is described by taking the limit $(t_{i+1} - t_i)=\Delta t_i\rightarrow 0$ for the discretized version of \Equ{eq:ItoSDE}:
\begin{align}
 x(t_{i+1}) = x(t_i) + a(x(t_i), t_i)\Delta t_i + b(x(t_i), t_i)\Delta W_i \label{eq:DiscreteSDE} \quad ,
\end{align}
where $\Delta W_i=W(t_{i+1})-W(t_i)$ and the grid points are time ordered $t_0\leq t_1 \leq \dots \leq t_n = t$.

\subsection[From FPE to SDE]{Connection between Fokker-Planck and Stochastic Differential Equations}
\label{ssec:FP2SDE}
The Fokker-Planck equation (FPE) is a parabolic partial differential equation and can be seen as a special case of the differential form of the Chapman-Kolmogorov equation with 
vanishing jump probability (e.g.\ \cite{gardiner2009}). In general, a FPE can be written as:
\begin{align}
 \frac{\partial n({\bf x}, t; {\bf y}, t')}{\partial t} = -\sum_i \frac{\partial}{\partial x_i}[A_i({\bf x}, t)n({\bf x}, t; {\bf y}, t')] + \frac{1}{2}\sum_{i,j}\frac{\partial^2}{\partial x_i \partial 
x_j}[B_{ij}({\bf x}, t)n({\bf x}, t; {\bf y}, t')] \label{eq:generalFPE} \quad , 
\end{align}
where $A_i({\bf x}, t)$ is the drift vector, $B_{ij}({\bf x}, t)$ is the diffusion tensor and $n({\bf x}, t; {\bf y}, t')$ is the density at place ${\bf x}$ and time $t$ depending on the density at place ${\bf y}$ 
and time $t'$. Here, ${\bf x}$ and ${\bf y}$ are in principle higher dimensional phase space vectors. The propagation of the phase space elements is described by the 
corresponding stochastic differential equation (here for the four dimensional case with three spatial and one momentum dimension):
\begin{align}
 {\rm d}r_\nu = A_\nu\,{\rm d}t+ D_{\nu\mu}\,{\rm d}\omega^\mu \label{eq:SDE} \quad ,
\end{align}
where ${\rm d}t$ is the time increment, $r_\nu$ is a 4-dimensional vector 
$(\vec{r}, ||\vec{p}||)$ and ${\rm d}\omega^\mu = \sqrt{{\rm d}t}\;\eta^\mu$ stands for a 4-dimensional 
Wiener process with Gaussian noise. The square-root proportionality of ${\rm d}\omega$ to ${\rm d}t$ is easily connected to the general behavior of normal diffusion $\langle x^2\rangle/t\propto D$. $A_\nu$ is a transport vector, $D_{\nu\mu}$ is a 4x4 tensor representing the diffusion and 
$\eta^\mu$ is a 4-vector of independent normal distributed random variables with zero mean and unity variance.

The derivation of \Equ{eq:SDE} from \Equ{eq:generalFPE} is beyond the scope of this paper and the interested reader is referred to textbooks like e.g.\ \cite{gardiner2009}. 
The connection between the FPE in \Equ{eq:generalFPE} and the SDE in \Equ{eq:SDE} can be compared to Lioville's equation and the ordinary differential equation which describes a 
deterministic motion:
\begin{align}
 \frac{\partial n({\bf x}, t; {\bf y}, t')}{\partial t} = -\sum_i \frac{\partial}{\partial x_i}[A_i({\bf x}, t)n({\bf x}, t; {\bf y}, t')], \quad \frac{{\rm d}x}{{\rm d}t} = A[{\bf x}(t)] \label{eq:Liouville} 
\quad .
\end{align}
Indeed Liouville's equation is only a special case of the Fokker-Planck equation for vanishing stochastic motion.


Since the spatial and momentum operators in \Equ{eq:DiffusionEquation} decouple for most Galactic applications, \Equ{eq:SDE} can be split into two parts:
\begin{align}
  {\rm d}\vec{r} &= \vec{A}\,{\rm d}t+ D_r\,{\rm d}\vec{\omega}_r \label{eq:SDE_spatial} \\
  {\rm d}q &= A_q\,{\rm d}t+ D_{qq}\,{\rm d}\omega_q \label{eq:SDE_momentum} \quad , 
\end{align}
where the absolute value of the momentum is denoted as $q=||\vec{p}||$, see e.g.\
\cite{Kopp2012} for a more detailed explanation. 

The remaining problem is the calculation of the tensor $D_r$ and scalar $D_q$ in the SDE from the given spatial diffusion tensor $\hat{\kappa}_{r}$ and the given momentum 
diffusion 
scalar $/\kappa_{qq}$, respectively, according to \Equ{eq:DiffusionEquation}. In general, one has to find the square root of the diffusion 
tensor $\hat{\kappa}$, which is the solution of $(\kappa+\kappa^t)=DD^\dagger$. In our case this reduces to a simple uncoupled set of equations, because we neglect 
drift terms and solve the equation in the local frame of the magnetic field line. Hence, the 
diffusion tensor is always diagonal which leads two the following relations:
\begin{align}
 D_{ij} = \delta_{ij} \sqrt{2\kappa_{ij}}, \qquad D_{qq} = \sqrt{2 \kappa_{qq}} \quad .
\end{align}

Note that while the SDE method allows for the computation of `quasi-trajectories' 
of particles and thereby, as mentioned above, makes it most compatible to the single particle tracking used for extragalactic cosmic rays, it nonetheless requires the knowledge of the diffusion tensor. This is opposite to so-called full-orbit simulations e.g.\ \citep{Giacalone-Jokipii-1999,Qin-Shalchi-2016}
where the elements of the diffusion tensor are determined from the particle trajectories obtained from a direct solution of their equations of motion.

\subsection{Numerical Implementation}
\label{ssec:Code}
In this section we will briefly explain the developed algorithm which calculates the next pseudo-particle position. First, we discuss the Euler-Mayurama scheme which can be used 
to integrate the SDE when the local orthonormal basis is known. In \Sec{sssec:Trihedron} the implemented adaptive method for the calculation of the local trihedron --- orthonormal 
basis defined by the Frenet equation --- is explained.

\subsubsection{The Euler-Maruyama Scheme}
\label{ssec:technical}
To explain the diffusion code all terms except spatial diffusion are neglected in this section.
Following \Equ{eq:SDE_momentum} and \Equ{eq:SDE_spatial} the so called 
Euler-Maruyama scheme (e.g.\ \cite{Desmond2001}) can be derived:
\begin{align}
 \vec{x}_{n+1} &= \vec{x}_n + D_r\,{\rm \Delta}\vec{\omega}_r \notag \\
	 &= \vec{x}_n + \left(\sqrt{2\kappa_{\parallel}}\,\eta_\parallel\, \vec{e}_t + 
\sqrt{2\kappa_{\perp, 1}}\,\eta_{\perp, 1}\, \vec{e}_n +
\sqrt{2\kappa_{\perp, 2}}\,\eta_{\perp, 2}\, \vec{e}_b\right) \sqrt{h} \label{eq:EMScheme}\quad ,
\end{align}
where $h=t_{i+1}-t_i$ is the integration time step.
Further, the orthonormal basis\footnote{The orthonormal basis is the local trihedron of the magnetic field line $\vec{B}(\vec{x}_n)$ when it is interpreted as a three-dimensional curve.} $\{\vec{e}_t, \vec{e}_n, \vec{e}_b \}$ is generally defined 
by the Frenet-Serret-equation (e.g.\ \cite{1969ArRMA..32...29M}) which can be rewritten for magnetic field lines (e.g.\ \cite{Effenberger2012}) as:
\begin{align}
 \vec{e}_t &= \vec{B}_{reg}/B_{reg} \notag \\ 
 \vec{e}_n &= (\vec{e}_t\cdot\nabla)\vec{e}_t/k \label{eq:trihedron}\\
 \vec{e}_b &= \vec{e}_t\times\vec{e_n} \notag \quad ,
\end{align}
where $\vec{B}_{reg}$ is the regular or coherent background field vector and $k=|{\rm d^2}\vec{r}/{\rm d}s^2|$ is the curvature of the parametrized field line 
$\vec{r}(s)$.\footnote{In this paper the term ``field line'' always refers 
to the 
coherent background field vector $\vec{B}_{reg}$.}. Generally, all three vectors of the local trihedron  $\{\vec{e}_t, \vec{e}_n, \vec{e}_b \}$ have to be calculated on the 
fly at each time step. Although 
the two perpendicular diffusion coefficients differ in principle, we state for the Galactic scenario that they can be assumed equal 
$\kappa_{\perp, 1} = \kappa_{\perp, 2}$. Anisotropic perpendicular diffusion can, in general, either result from anisotropic (or non-axisymmetric) turbulence perpendicular to the mean magnetic field \cite{Ruffolo_2008}, which is 
not the case for the Galactic magnetic field (see the turbulent component of the JF12 field \cite{JAN12}). Otherwise the decoupling of the perpendicular directions is caused by a 
large curvature of the background field on scales of the mean scattering length of the particles. This is also not observed for the Galactic magnetic field. This is in contrast 
to simulations in the heliosphere, where the consideration of anisotrpic perpendicular diffusion can be crucial (see e.g.\ \cite{EFF11}).

\subsubsection{Adaptive Calculation of the Local Trihedron}
\label{sssec:Trihedron}
The tangential vector of the magnetic field line $\vec{e}_t$ is calculated via the so called 
Cash-Karp (CK) algorithm \cite{CAS90}, an adaptive Runge-Kutta (RK) algorithm. With this algorithm we are able to adapt the time integration step to minimize the number of steps. 
The local truncation error for the field line integration can be set by the user. In this way the overall computation time is reduced without any losses in accuracy. 

The tangential vector is approximated by the difference of two points on the magnetic field line $\vec{e}_t=\vec{r}_{end}-\vec{r}_{start}$. Here, the starting point is given by 
the old pseudo-particle position $\vec{r}_{start} = \vec{x}_n$ (see \Equ{eq:EMScheme}) and the end position $\vec{r}_{end}$ is calculated with a field line integration:
\begin{align}
 \vec{r}_{end} = \vec{r}_{start} + \int_0^L \vec{B}/B\,{\rm d}s \label{eq:FieldLineIntegration}\quad ,
\end{align}
where $L=\sqrt{2\kappa_{\parallel}}\,\eta_\parallel\,\sqrt{h}$ is the arc length of the field line elements and has the length of the diffusion step in parallel direction. The CK 
algorithm 
uses a fourth and fifth order RK algorithm to calculate two solutions of the initial condition problem given in \Equ{eq:FieldLineIntegration}. The norm of the 
difference of the vectors is used as a measure for the local truncation error. The step is accepted if:
\begin{align}
 m = ||\vec{r}_{end, 4}-\vec{r}_{end, 5}|| \leq \xi\cdot{\rm kpc} \label{eq:TruncError} \quad ,
\end{align}
where $\xi$ is the user set precision and $\vec{r}_{end, 4}$ and $\vec{r}_{end, 5}$ are the $4^{th}$- and $5^{th}$-order solution, respectively.

In the case of a rejected step integration length $L$ is bisected until the condition in \Equ{eq:TruncError} is fulfilled or the minimum step is reached. Following this procedure 
the end position vector $\vec{r}_{end}$, used for the calculation of the tangential vector $\vec{e}_t$, is derived by $2^n$ consecutive solutions of \Equ{eq:FieldLineIntegration} 
if ($n-1$) step attempts are rejected:
\begin{align}
 \vec{r}_{end} = \vec{r}_{start} + \sum_{j=0}^{2^n-1} \; \int_{L_j}^{L_{j+1}} \vec{v}(s) \,{\rm d} s \label{eq:RejectionSum} \quad ,
\end{align}
where $L_j=2^{-n}L\,j$ and $\vec{v}=\vec{B}/|\vec{B}|$ is the normalized magnetic field direction. This procedure is not 
as efficient as the conventional step adaption ($h_{next}=0.95\,h\cdot m^{-0.2}$ see \cite{CAS90}) but is necessary. If the step size is 
reduced as conventional, large values of $\eta_\parallel$ will be rejected more frequently as compared to small values. This would lead to an effective underestimation of the 
tails of the Wiener process ${\rm d}\omega$. Furthermore, this would not just be an inaccurate description of the diffusion process but is simply not an approximation of the 
diffusion. The error is hard to quantify since it depends on the global magnetic field structure, the diffusion tensor and also on the simulation set-up such as 
minimum/maximum step size and the precision. With the procedure of \Equ{eq:RejectionSum} we ensure that the particles are always 
propagated over the full drawn diffusion step but use an adequate number of intermediate steps. 

The suggested next integration time $h_{next}$ is:
\begin{align}
 h_{next} &= h \cdot 2^{(-2n)} \quad \textrm{if}\; n\geq 0\\
  h_{next} &= h \cdot 4 \qquad\quad \textrm{else} \quad .
\end{align}
When the current step is not decreased the next step should be increased. The next time ingration time step $H_{next}$ is increased by a factor 4, which leads on average to a 
doubled parallel step length $L$, due to the square root dependence of the step length from the integration time.

%% file: Validation.tex
\section{Validation of the Algorithm}
\label{sec:Validation}

The new code is validated in two ways. Firstly, we proof that our code reproduces the 
correct solution for a given diffusion coefficient. In doing so, we compare in \Sec{ssec:Analytical} the 
simulation with simple analytical expectations. Further, we validate the adaptive field line 
integration in section \Sec{ssec:FieldLineIntegration}.

All tests that are developed for this paper test a mathematical problem. That means they are not designed to describe the physical reality. So before the software is used for physical simulations it should be carefully examined if the necessary assumptions made for the tests do also hold (at least approximately) in the physical simulation.

\subsection{Homogeneous Background Field}
\label{ssec:Analytical}

\subsubsection{Green's function}
For the first test we simulate protons under the influence of diffusion in a homogeneous 
magnetic field $\vec{B}=B_0\cdot\vec{e}_z$. We use a diffusion tensor that is anisotropic with 
respect to the background field $\hat{\kappa}={\rm diag}(\kappa_{\perp}, \kappa_{\perp}, 
\kappa_{\parallel})$ with $\kappa_{\parallel}=10\kappa_{\perp}$. Here, the probability 
distribution function (PDF or density $n$) of a pseudo 
particle is known 
and given by the Green's function of the following equation:
\begin{align}
 \frac{\partial n}{\partial t} &= \nabla \cdot (\hat{\kappa} \nabla n) + \delta({\vec{r}})\delta({t})
\quad ,
\end{align}
which is solved by
\begin{align}
 n(r_i, t) = \frac{1}{(4\pi \kappa_{ii}t)^{d/2}}\cdot\exp\left(-\frac{r_i^2}{4\kappa_{ii}t}\right)
\quad .
\label{eq:Dif_Solution}
\end{align}
Here, $\kappa_{ii}$ is the diffusion coefficient, $d$ is the dimension ($d=1$ in this test case, 
because the three directions are decoupled) and $r_i$ is the $i^{\rm th}$ component 
of the spatial vector. In other words, we expect that the distributions of the three vector components ($x(t), y(t), z(t)$) of the pseudo-particles follow normal distributions with zero means and variances $\sigma_i=\sqrt{2\kappa_{ii}t}$ since that is the analytical solution as given in \Equ{eq:Dif_Solution}.

\begin{figure}
 \includegraphics[width=\textwidth]{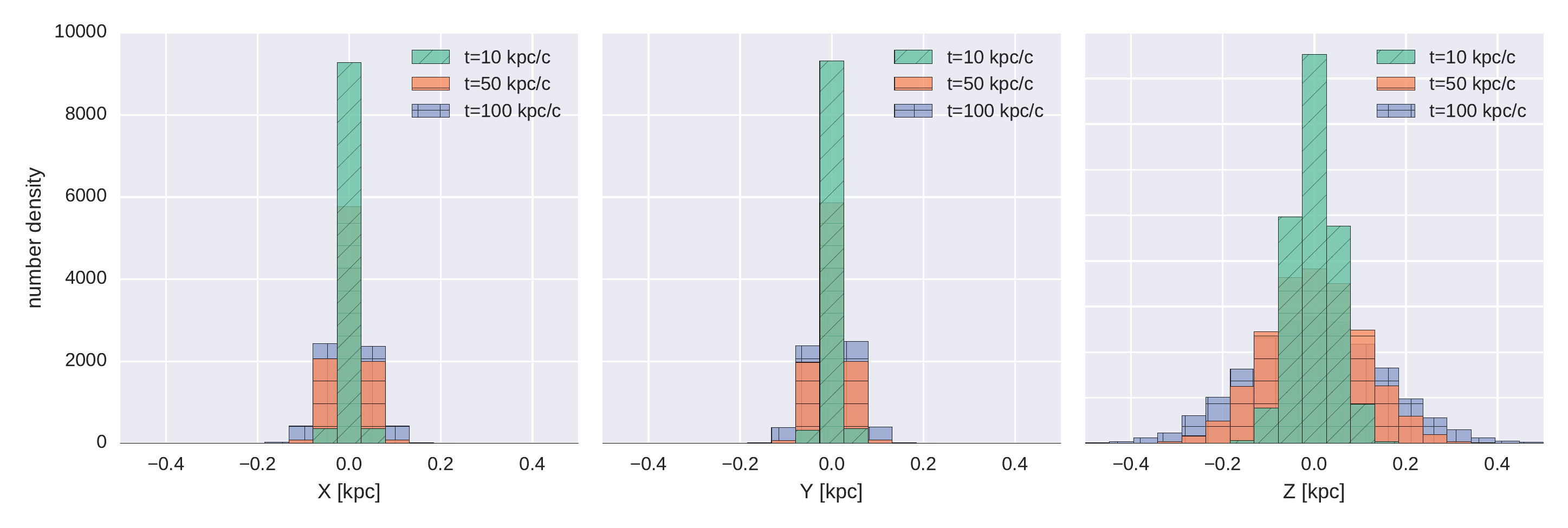}
 \caption{Diffusion in a homogeneous magnetic field for different times. The magnetic field is 
aligned with the z-axis.}
 \label{fig:Diffusion1}
\end{figure}

\begin{figure}
 \includegraphics[width=\textwidth]{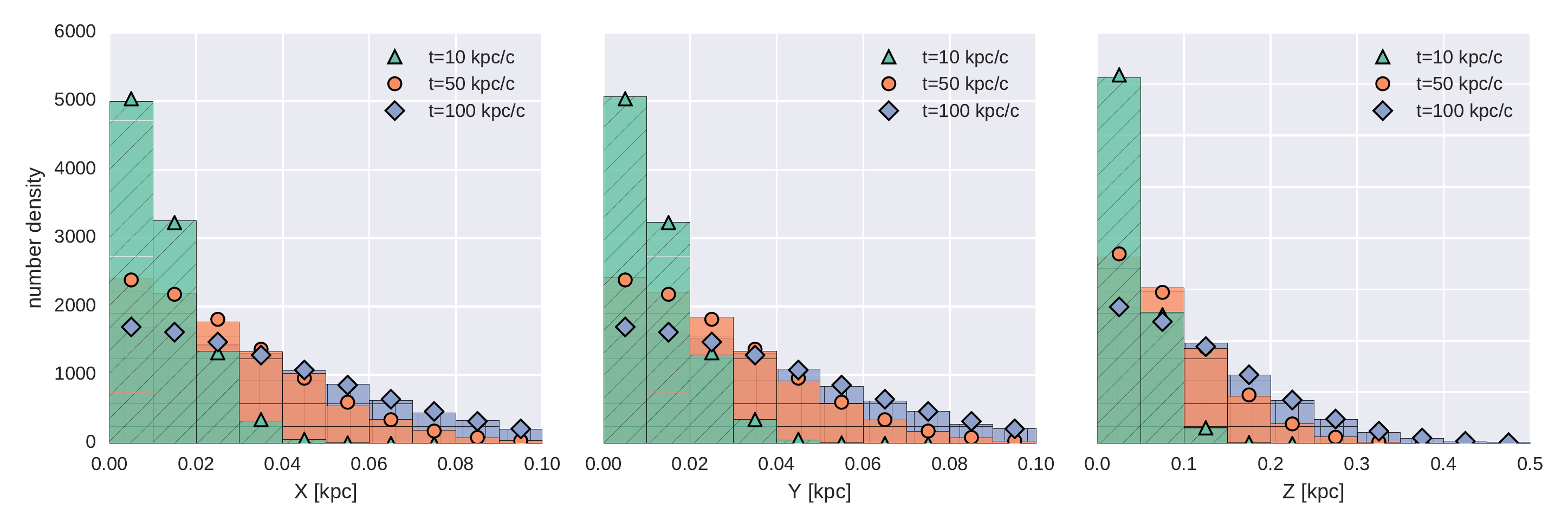}
 \caption{Diffusion in a homogeneous magnetic field for different times. The distance from the 
origin is shown for all three components. The simulated data are shown as a histogram 
and the analytical expectations as lines.}
 \label{fig:Diffusion2}
\end{figure}

\Figure{fig:Diffusion1} shows the density distributions for all three components at three different 
times. The difference between the parallel (z-) and the perpendicular 
(x-, y-) direction is clearly visible. Furthermore, it can be seen that the width of the distribution increases with increasing time and that the speed of this process depends on 
the diffusion coefficient. To emphasize this diffusive behavior we plot the absolute distance from the origin for all three components in combination with the analytical 
solution in \Fig{fig:Diffusion2}. A difference between the analytical solution (line) and the 
simulated data (histogram) cannot be seen by eye. To quantify these 
results a statistical test is applied to the data.

Since we expect that the differences per bin between the simulated and expected distribution are normally distributed we use a  $\chi^2$-test of the probability distribution function (PDF) of the particle range (norm of the end position vector 
components). So here, $P(r)$ is tested rather than $P(\vec{r})$. This PDF is described by:
\begin{align}
  \tilde{P}(R_i) &= 
\frac{2}{\sqrt{4\pi\kappa_{ii}t}}\cdot\exp{\left(-\frac{R_i^2}{4\kappa_{ii}t}\right)} 
\label{eq:RadialPDF} \quad .
\end{align}
Here, $R_i=|r_i|$ is the absolute value of the vector components. 
The distance is sorted into bins while it is ensured that the minimal count per bin is greater
than 5. This is done by a consecutive reduction of the number of bins. These bin counts are 
compared to the expected values derived from the 
integration of \Equ{eq:RadialPDF} over the bin edges. From that the $\chi^2$ value is 
calculated as:
\begin{align}
 \chi^2 = \sum_{i=1}^n \frac{(O_i-E_i)^2}{E_i} \label{eq:Chi2}\quad ,
\end{align}
where $n$ denotes the number of bins and $O_i$ and $E_i$ the observed and expected count of 
events in the $i^{th}$ bin, respectively. 
Using the $\chi^2$ test result according to \Equ{eq:Chi2} we calculate the p-value:
\begin{align}
p(\chi^2) = \int_{\chi^2}^\infty P_k(\chi^2)\;{\mathrm d} \chi^2 \label{eq:pValue} \quad , 
\end{align}
where $p$ is the p-value\footnote{Although the p-value is seen as deprecated by the American Statistical Association among others it is not critical to use it here.} and $P_k(\chi^2)$ is the $\chi^2$-distribution with $k$ degrees of freedom.

The Null hypothesis, that the sample is drawn 
from \Equ{eq:RadialPDF}, is rejected if the p-value is not in the range ($0.005<p<0.995$). We choose this broad range of acceptance because we want to minimize the number of false rejections. Since problems with the algorithm, at least to our experience, lead to very large deviations we decided to use this robust acceptance range. This also allows us to use this test for the automated testing of CRPropa (see below). Here, a large number of statistical tests is applied and the test suite has to be rerun if one of them fails, so a very robust test setup is desired.

According to the p-values summarized in \Tab{tab:ChiSquareSimpleDif}, we conclude that the 
algorithm reproduces the diffusive behavior correctly in the case of a homogeneous background field 
using an anisotropic diffusion 
tensor.
\begin{table}[htbp]
\caption{P-values for diffusion in homogeneous background field}
\centering
\begin{tabular}{c|ccc}
\toprule
\diagbox{Coordinate}{Time [s]} & $10^{12}$ & $5.2\cdot10^{12}$ & $10^{13}$ \\
\midrule
X & 0.64 & 0.95 & 0.06 \\
Y & 0.94 & 0.85 & 0.02 \\ 
Z & 0.23 & 0.58 & 0.30 \\ 
\bottomrule
\end{tabular}
\label{tab:ChiSquareSimpleDif}
\end{table}
\begin{table}[htbp]
\caption{$\chi^2$/dof for diffusion in homogeneous background field}
\centering
\begin{tabular}{c|ccc}
\toprule
\diagbox{Coordinate}{Time [s]} & $10^{12}$ & $5.2\cdot10^{12}$ & $10^{13}$ \\
\midrule
X & 0.78 & 0.37 & 1.80 \\
Y & 0.38 & 0.48 & 2.22 \\ 
Z & 1.29 & 0.84 & 1.19 \\ 
\bottomrule
\end{tabular}
\label{tab:ChiSquareSimpleDif2}
\end{table}

\paragraph{Unittest in CRPropa 3.1}
A slightly different version of the tests described above is implemented in the unit 
test framework of CRPropa\,3.1 which can be run within the installation process of the 
software. In this way it is ensured that the code is executed as expected. If all tests are passed, 
the user will be sure that no known software problems occur after the 
installation.

Here, two statistical tests are implemented. First, the Anderson-Darling test
\cite{Anderson1952} is applied on the simulated data. This test can be compared to the commonly known Kolmogorov-Smirnov test but gives more weight to the tails of the 
distribution. The Anderson-Darling test is very sensitive to outliers of the data but cannot be used to test against a specific normal distribution. The Anderson-Darling turned out to be a very good test to find problems in the numerical implementation of the diffusion algorithm. Since is it sensitive to outliers the test detects even a small number of failed pseudo-particle propagations which was helpful during the development of the code. 

If the Anderson-Darling test does not find any significant deviations from a normal distribution we can assume that the errors of the simulated data compared with the analytic results will follow also a normal distribution. This allows us to apply the $\chi^2$-test to verify that the width of the normal distribution does not deviate significantly from the expected one.

\subsubsection{Stationary Test Solution}
\label{sssec:PICARD}
This test, as the one before, is not a physical model but rather a artificial problem specifically designed to test the algorithm. Therefore a comparison with real physical problems can be misleading.

In this section the test setup for a more complex simulation which is also used in the 
testing of the PICARD code \cite{Kissmann2014} is explained.
Here, the stationary solution of a simple diffusion equation:
\begin{align}
 -\nabla\cdot(\hat{\kappa} \nabla n(\vec{r})) &= s(\vec{r}) \quad , \label{eq:PICARDTest}
\end{align}
with $\hat{\kappa} = {\rm diag}(\kappa_{xx}, \kappa_{xx}, \kappa_{zz})$  and a source term  $s(\vec{r})$ is tested against the known analytical solution. With the boundary 
condition $n(x=\pm R, y=\pm R, z=\pm H)=0$ and the source term defined by
\begin{align}
s(\vec{r}) = 
\frac{4}{\pi^2} \left(\frac{2\kappa_{xx}}{4R^2} + \frac{\kappa_{zz}}{2H^2} \right) \cdot
\cos\left(\frac{x\pi}{2R}\right)\cos\left(\frac{y\pi}{2R}\right)\cos\left(\frac{z\pi}{2H}\right)  
\label{eq:PICARD_source} \quad ,
\end{align}
where the solution is given by:
\begin{align}
n_{ana}(\vec{r}) = 
\cos\left(\frac{x\pi}{2R}\right)\cos\left(\frac{y\pi}{2R}\right)\cos\left(\frac{z\pi}{2H}\right) 
\quad .
\label{eq:PICARD_analyticSolution}
\end{align}

For the test procedure we inject pseudo-particles using a simple rejection sampling method. In that way we make sure that the source distribution follows \Equ{eq:PICARD_source} 
ignoring the constant factor for now. 
Afterward they are propagated forward in time and the position of the particles is recorded 
at $N$ 
consecutive times $t_1, t_2, ..., t_N$. The boundary condition is implemented simply by deactivating the candidates when they reach the boundary. This removal of the pseudo-particles from the simulation leads automatically to $n(x=\pm R, y=\pm R, z=\pm H)=0$ as required.

The time dependent particle distribution $n(t_i)$ 
corresponds to the solution of the following equation:
\begin{align}
 \frac{\partial n}{\partial t} = \nabla\cdot(\hat{\kappa} \nabla n) + s(\vec{r})\delta(t-t_i) \label{eq:DeltaDiffusion} \quad ,
\end{align}
which differs from \Equ{eq:PICARDTest} since it is not a stationary equation. We assume that the 
particle distribution 
$n(t_i)$ does not vary significantly in the time interval $h_i = t_{i+1}-t_i$ so we can calculate the 
stationary solution from our particle distribution as follows:
\begin{align}
 n_{sim}(\vec{r}) = \sum_{i=1}^{N_{snap}} n(t_i) h_i w \quad \label{eq:ApproxStationarySolution}.
\end{align}
Here, $w$ is a weight according to the source function, where $s(\vec{r})=w\cdot n_{ana}(\vec{r})$. In 
general, \Equ{eq:ApproxStationarySolution} is an infinite sum but the density $n(t)$ decreases 
with time $\lim_{t\rightarrow\infty} n(t)=0$. It follows that there exists a time $t=T_{\rm max}$ 
with $\int_{V} n(T_{\rm max})\,{\rm d}V\leq\beta N$ where only a mere fraction $\beta$ of the particle remain in the simulation volume. The maximum integration time $T_{\rm max}$ 
needs to be chosen such that $\beta$ is small enough for the aspired accuracy. In other words, since the time integration of \Equ{eq:DeltaDiffusion} leads to \Equ{eq:PICARDTest}, 
the discrete solution in \Equ{eq:ApproxStationarySolution} is an approximation for the solution of the original stationary transport equation.

This test is designed to answer several questions:
\begin{enumerate}
 \item How many snapshots $N_{\rm snap}$ are needed to resolve the solution sufficiently? The number of snapshots $N_{snap}$ determines the number of summands in 
\Equ{eq:ApproxStationarySolution}.
 \item What is a reasonable size of the integration time step $h$?
 \item What is the integration time $T_{\rm max}$ to reach the stationary solution?
\end{enumerate}

In this test we implemented the parameters as $\kappa_{zz}=10^{24}\,\rm{m^2/s}$, $\kappa_{xx}=0.1 
\kappa_{zz}$, $R=H=0.5\,\rm{kpc}$, $T_{\rm max}=50\,\rm{kpc/c}$ and $N=10^6$.
We used two different approaches: In the first one we fixed the integration time step 
$h=1\,{\rm pc/c}$ and varied the number of snapshots $N_{\rm snap}$. To answer the second 
question we fixed the number of snapshots $N_{\rm snap}=500$ and varied the integration time step 
$h$. In both setups we analyzed the integrated 
number density depending on the maximum integration time (see paragraph `Total particle number'). 
After that we examined the accuracy of the spatial density distribution 
$n_{sim}(\vec{r})$ for the latest time point $T_{max}=50\,\rm{kpc/c}$ (see paragraph `Spatial accuracy').

\begin{figure}
 \includegraphics[width=\textwidth]{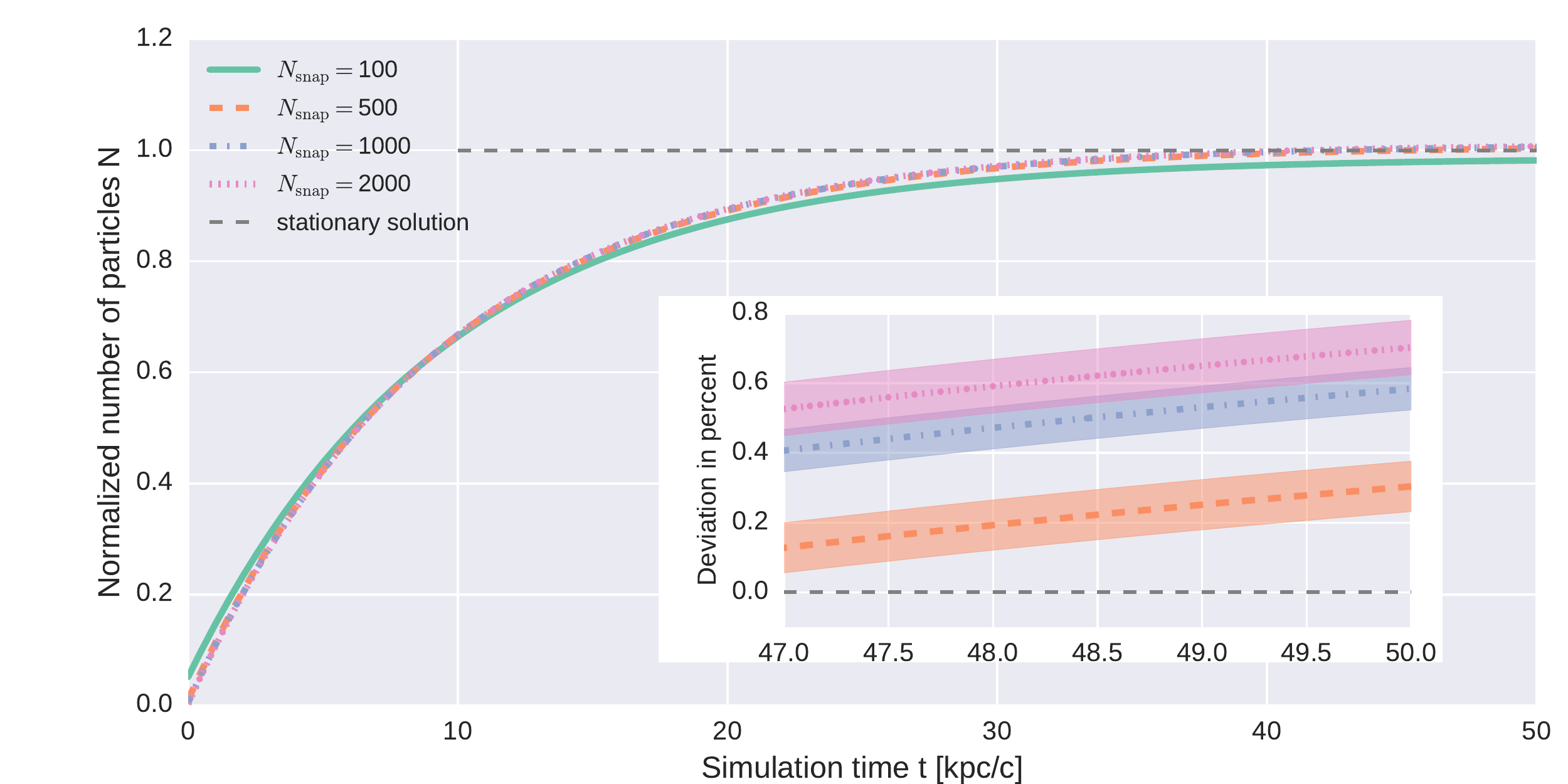}
 \caption{Integrated number density depending on the maximum integration time. Different colors show 
different numbers of snapshots $N_{snap}$. The shaded color bands correspond to a statistical 
uncertainty of one standard deviation.}
 \label{fig:PICARD_integrated_fixedStep}
\end{figure}

\begin{figure}
 \includegraphics[width=\textwidth]{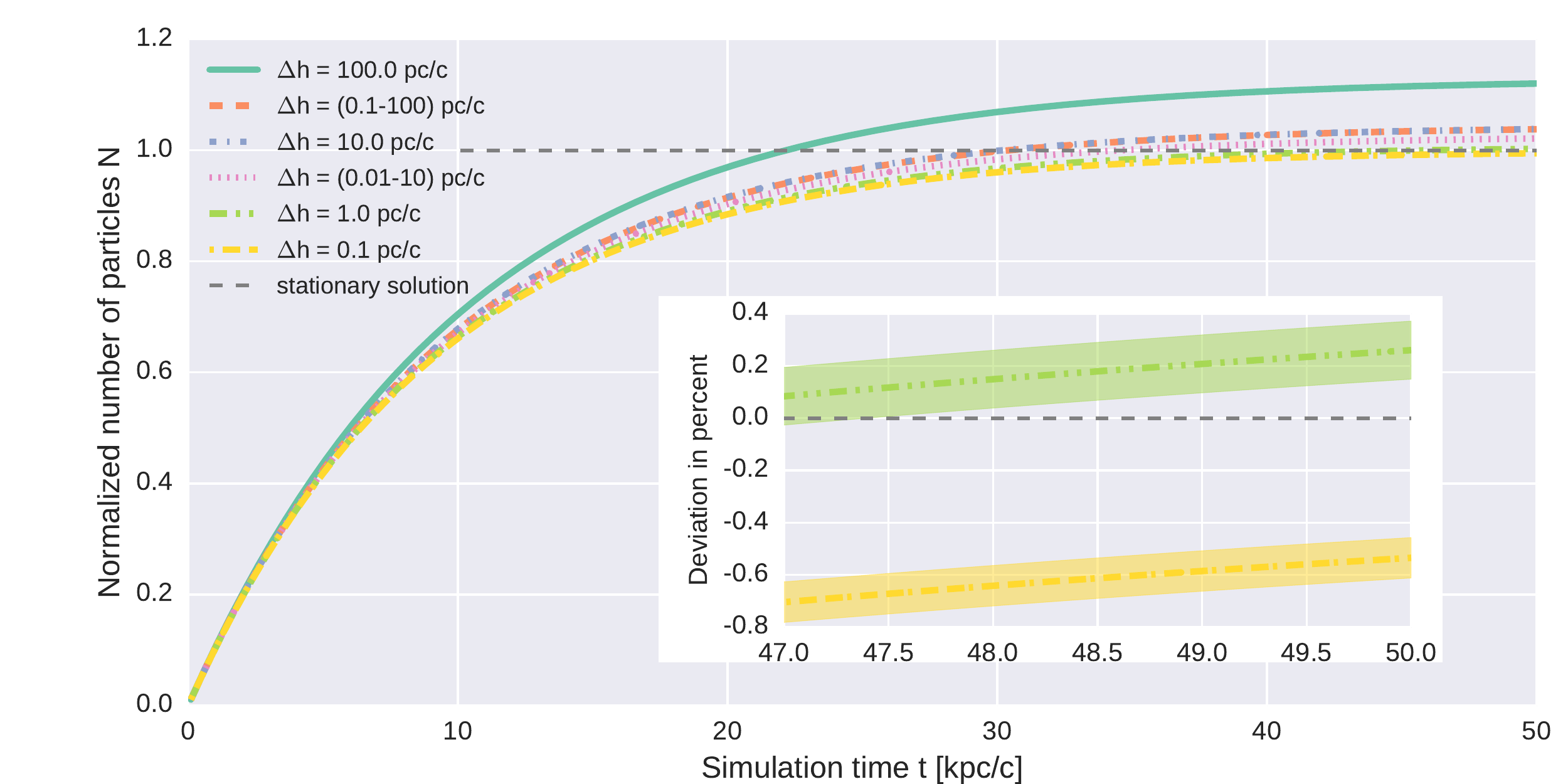}
 \caption{Integrated number density depending on the maximum integration time. Different colors show different integration time steps $h$. The shaded color bands correspond to one 
standard deviation statistical uncertainty.}
 \label{fig:PICARD_integrated_fixedSnap}
\end{figure}

\paragraph{Total Particle Number}
\Figure{fig:PICARD_integrated_fixedStep} and \Fig{fig:PICARD_integrated_fixedSnap} each show the 
simulated integrated particle density $\int_V n_{sim}(\vec{r})\,{\rm d}V$ depending on the 
maximum integration time. In general both figures show that the approximated solution asymptotically 
increases to a value close to the expectation.

\Figure{fig:PICARD_integrated_fixedStep} shows that the estimated density increases with an 
increasing number of snapshots. This is reasonable because a higher number of snapshots 
approximates the time evolution better. This gives effectively more weight to the early time points with higher densities as compared to the case with a low time resolution, 
where the first time snapshot is taken when already a significant number of particles is lost. The error for a 
sufficient number of snapshots $(N>500)$ is below one percent for $T_{\rm 
max}>50\,\rm{kpc/c}$. Furthermore, it is visible that the integrated particle number does not 
converge to a completely flat distribution, due to the fact that $\beta>0$. That means that not all 
pseudo-particles have time to leave the simulation volume before the end of the simulation time. A longer time 
integration can fix this error but is not computing-time efficient. Moreover, the density is slightly 
over-estimated. This is presumably caused by problems of the numerical implementation of the boundary conditions. To 
estimate the statistical error we repeated the simulation ten times and calculated the mean and standard deviation from these sets.

\Figure{fig:PICARD_integrated_fixedSnap} shows the results for different integration time steps $h$. Here, an increased accuracy (decreased time step) leads to a lower 
integrated particle number, where $h=0.1\,\rm{pc/c}$ gives an approximation smaller than the 
expected one. From that we conclude that an integration time step of  $h=0.1\,\rm{pc/c}$ is 
small enough to fully resolve the boundary. In addition, we allowed the adaptive algorithm to 
choose the step size depending on the pseudo-particle 
position. In doing so, the algorithm becomes significantly faster than it would have been using 
the minimal allowed integration step only. On the other hand, it is a lot more 
accurate than the solution for the maximum allowed step.

\paragraph{Spatial Accuracy}

After that the accuracy of the spatial density distribution 
$n_{sim}(\vec{r})$ is examined for 
$T_{max}=50\,\rm{kpc/c}$. In doing so, the data are binned in a three-dimensional histogram with 
ten 
bins in each direction to ensure sufficient statistics in each bin. Furthermore, the expected 
number of particles per bin is calculated using the bin edges and the analytic solution given in 
\Equ{eq:PICARD_analyticSolution}.

\begin{figure}
\centering
\begin{minipage}{0.5\textwidth}
\subfigure[][Binning and magnetic field direction (blue arrows)]{
\includegraphics[width = .9\textwidth]{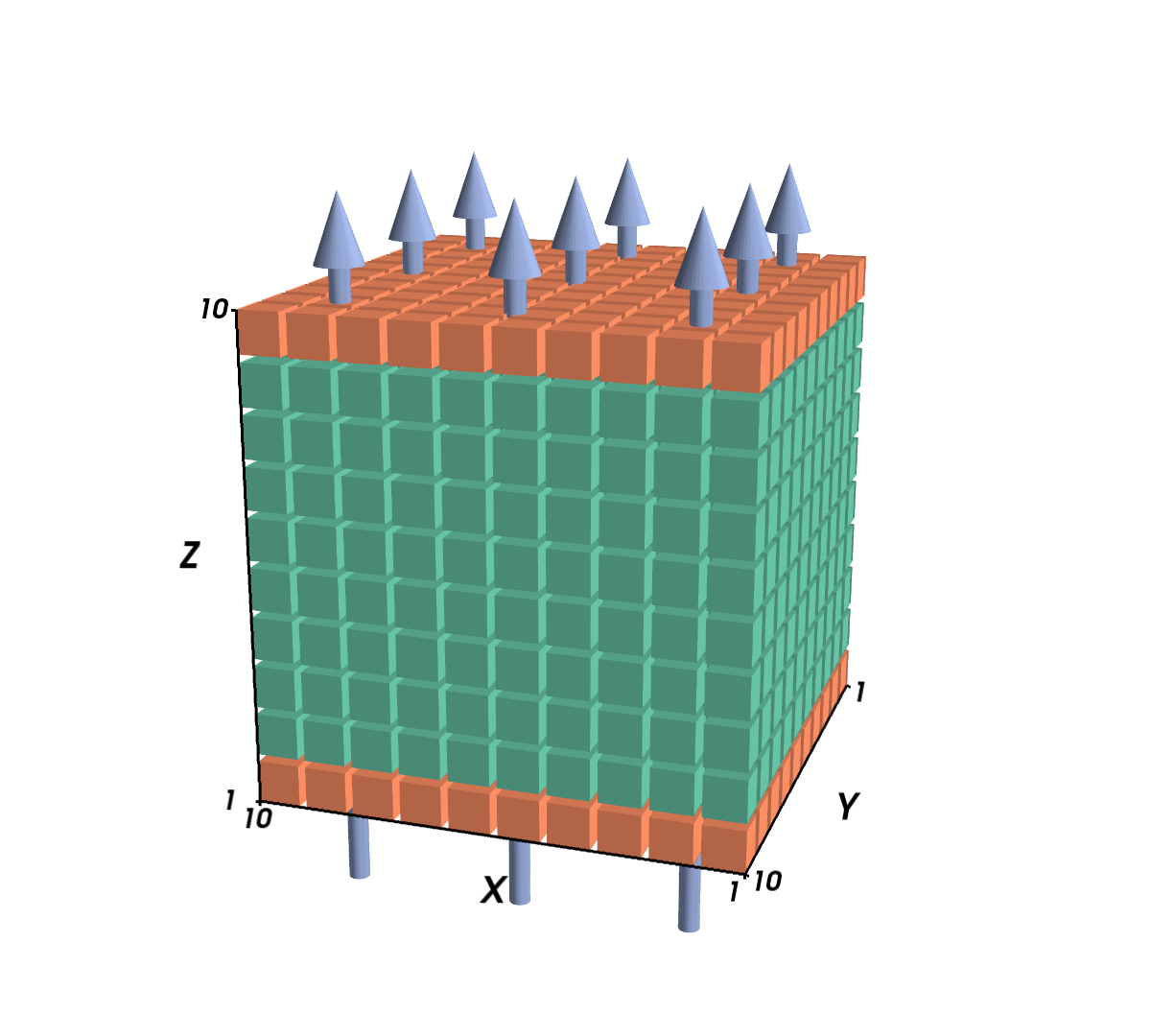}
\label{fig:Binning}
}
\end{minipage}%
\begin{minipage}{0.5\textwidth}
\subfigure[][Simulated approximation of the stationary density distribution. The density $n$ is colorcoded.]{
\includegraphics[width =  .9\textwidth]{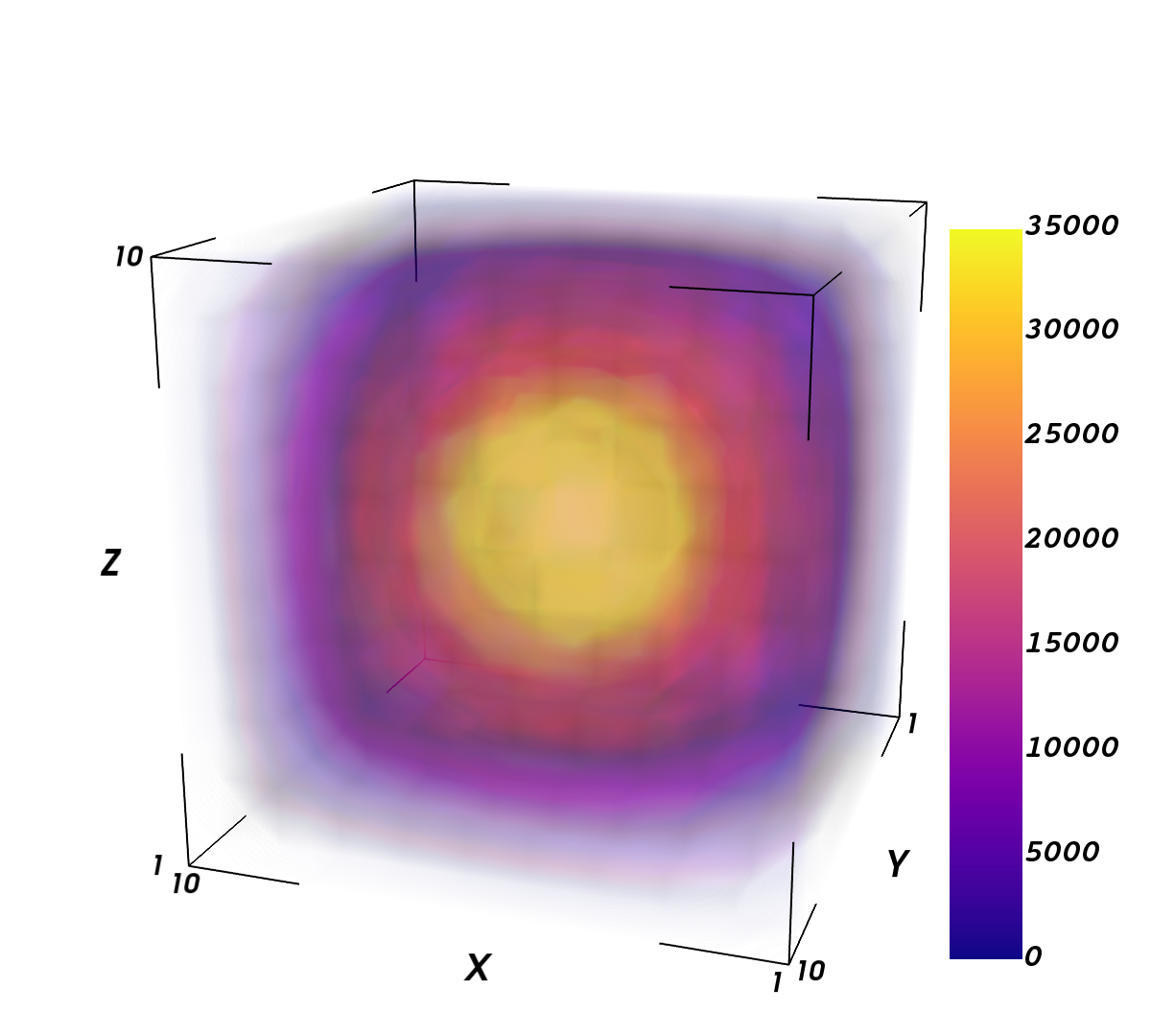}
\label{fig:Density3Dim}
}
\end{minipage}
\caption{In figure (a) the binning and the background field of the analysis are shown. Different colors refer to the anomaly of the mean error per bin because of the anisotropic 
diffusion tensor. The other figure (b) shows exemplary the time integrated density distribution \Equ{eq:ApproxStationarySolution} for 
$\Delta h=0.1\,\rm{pc/c}$ and $N_{snap}=500$. Here, all snapshots up to the maximum integration time $T_{max}=50\,{\rm kpc/c}$ are averaged to yield the most accurate 
approximation.}
\label{fig:SpatialAccuracy}
\end{figure}

A first analysis showed that the error depends on the bin position. More precisely we can identify two groups of bins: The two outer layers in z-direction (orange bins in 
\Fig{fig:Binning}) and the other bins (green bins in \Fig{fig:Density3Dim}). This anomaly of the errors is an effect of the anisotropic diffusion tensor. Since the mean spatial 
step is larger in z-direction than in the x-y-plane the boundary condition is resolved worse in 
these bins. In general, all simulations could reproduce the shape of the analytical 
solution. \Figure{fig:Density3Dim} displays exemplary that the density distribution for 
$N_{snap}=500$ and $\Delta h=0.1\,\rm{pc/c}$ reproduces the expected shape of 
\Equ{eq:PICARD_analyticSolution}.

\Table{tab:spatialErrorFixedStep} shows the relative errors of setup 1 (fixed integration step) 
broken down for the different bin groups. The table gives also the overall mean per 
simulation setup. Evidently, the increase from 1000 to 2000 snapshots has nearly no effect on the 
results but it doubles the amount of data being processed. Consequently, a good guess for the 
required accuracy for the problem saves a lot of memory and processing time. The decreasing accuracy 
from 500 to 1000 snapshots can be explained by 
the fact that the approximated solution is not yet fully stationary, meaning the maximum integration 
time $T_{max}$ is not long enough.
\begin{table}[htbp]
\caption{Mean relative error for different simulation setups with fixed integration time 
$h=0.1\,{\rm pc/c}$.}
\centering
\begin{tabular}{c|cccc}
\toprule
\diagbox{Bins}{$N_{snap}$} & 100 & 500 & 1000 & 2000 \\
\midrule
Z-boundary & 0.023 & 0.045 & 0.048 & 0.049 \\ 
Inner & -0.019 & 0.002 & 0.006 & 0.006 \\ 
All & -0.011 & 0.011 & 0.014 & 0.015 \\ 
\bottomrule
\end{tabular}
\label{tab:spatialErrorFixedStep}
\end{table}

\Table{tab:spatialErrorFixedSnap} clearly exposes that a decreased integration time step $h$ 
decreases the relative error effectively. The good performance of the adaptive 
algorithm is confirmed also in the spatial analysis. The algorithm uses 
larger time steps in the middle of the simulation and decreases the integration time at the 
boundary. This results in a better accuracy than reached using the upper 
integration time limit at low computational costs. However, the accuracy of the lower time integration limit is not met.
\begin{table}[htbp]
\caption{Mean relative error for different simulation setups with fixed number of snapshots 
$N_{snap}=500$.}
\centering
\begin{tabular}{c|cccccc}
\toprule
\diagbox{Bins}{Time step} & 100 & 0.1-100 & 10 & 0.01-10 & 1 & 0.1 \\
\midrule
Z-boundary & 0.554 & 0.188 & 0.172 & 0.12 & 0,045 & 0.008 \\ 
Inner & 0.113 & 0.033 & 0.036 & 0.019 & 0.002 & -0.005 \\ 
All & 0.201 & 0.064 & 0.063 & 0.039 & 0.01 & -0.003 \\ 
\bottomrule
\end{tabular}
\label{tab:spatialErrorFixedSnap}
\end{table}

A more detailed analysis, including full error resolution per bin and spread of the error in the 
different groups, is given in appendix \ref{ap:SpatialError}.

\subsection{Field Line Integration}
\label{ssec:FieldLineIntegration}
As it is shown in \Sec{ssec:Analytical} the code reproduces the solution of the transport equation 
for a homogeneous magnetic field very well. The next step is to check whether the field line 
integration or derivation of the tangent vector $\vec{e}_t$ works. Therefore, we developed two 
different tests. The first problem has a simple analytic solution which enables us 
to calculate errors very carefully (see \Sec{sssec:SpiralIntegration}). The second one uses a 
realistic magnetic field as it may be chosen for a full simulation setup but 
does not provide easy analytic solutions (see \Sec{sssec:JF12FieldlineIntegration}).

As explained in \Sec{ssec:technical} the algorithm is designed to adapt the integration 
time to follow the magnetic field line with the user defined precision. The basic assumption 
of this test is pure parallel diffusion, i.\ e.\ $\kappa_\perp=0$. Since magnetic field lines do 
not intersect, particles cannot leave their original field line. We make use of this fact 
in the 
following tests as we use the deviation from the field line as a measure for the accuracy of the 
algorithm.

\subsubsection{Spiral Line}
\label{sssec:SpiralIntegration}
\begin{figure}
\centering
 \includegraphics[height=.35\textheight]{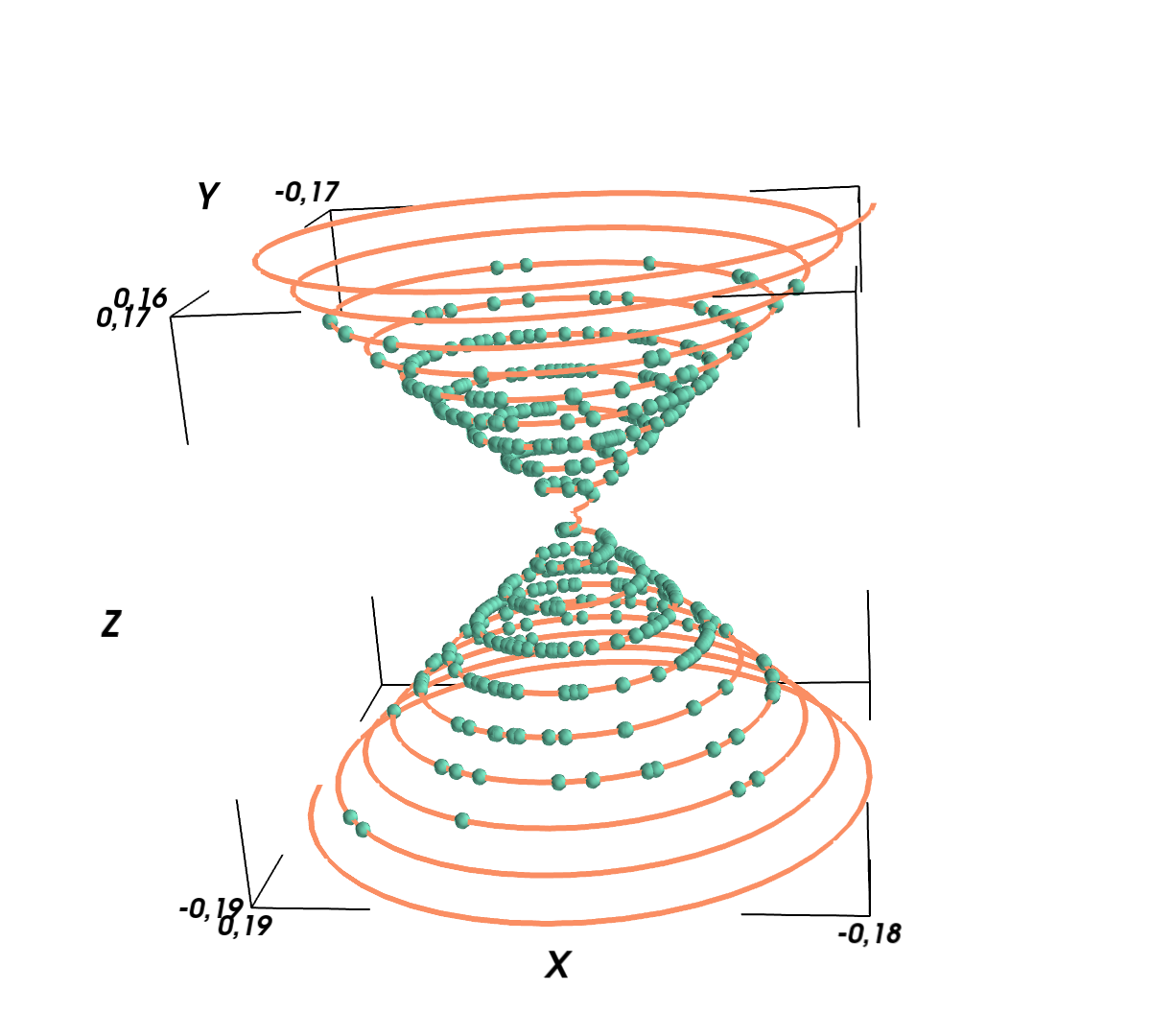}
 \caption{End positions of 500 pseudo-particles (blue spheres) and field line (orange line). 
Particles were injected at the middle of the double spiral at $\vec{r}_0=\vec{0}$.}
 \label{fig:SpiralExample}
\end{figure}

In this section we explain the test of the field line integration of the developed 
algorithm. A simple curve $\vec{r}_{\rm spiral}$ of the form:
\begin{align}
  \vec{r}_{\rm spiral}(z) = 
  \begin{pmatrix} 
    z\cdot\cos(2\pi z/s) \\ 
    z\cdot\sin(2\pi z/s) \\ 
    z 
  \end{pmatrix} \label{eq:Spiral}
\end{align}
is used as the field line. The curve is parametrized in $z$ and the constant $s$ 
determines the windings per height. The form of the field line is shown in \Fig{fig:SpiralExample}. 
Due to the varying curvature radius of the spiral \Equ{eq:Spiral} this curve is ideal to test the 
adaptive algorithm.
In doing so, we inject 10{,}000 particles at the origin $\vec{r}_0=\vec{0}$ into the simulation 
and propagate them forward for a time span of $T_{max}=100\,{\rm kpc/c}$. We used a minimum/maximum integration 
step of $h_{\rm min}=10^{-5}\,{\rm kpc/c}$ and $h_{\rm max}=1\,{\rm kpc/c}$, respectively. In different simulations the local truncation error $\xi$ is varied. The 
diffusion coefficient in parallel direction is fixed to $\kappa_\parallel\approx 6.4\cdot 10^{25}\,{\rm m^2/s}$ and the diffusion in perpendicular direction is neglected 
$\kappa_\perp=0$. The winding constant is fixed at $s=0.02\,{\rm kpc}$. 

\begin{figure}
 \includegraphics[width=\textwidth]{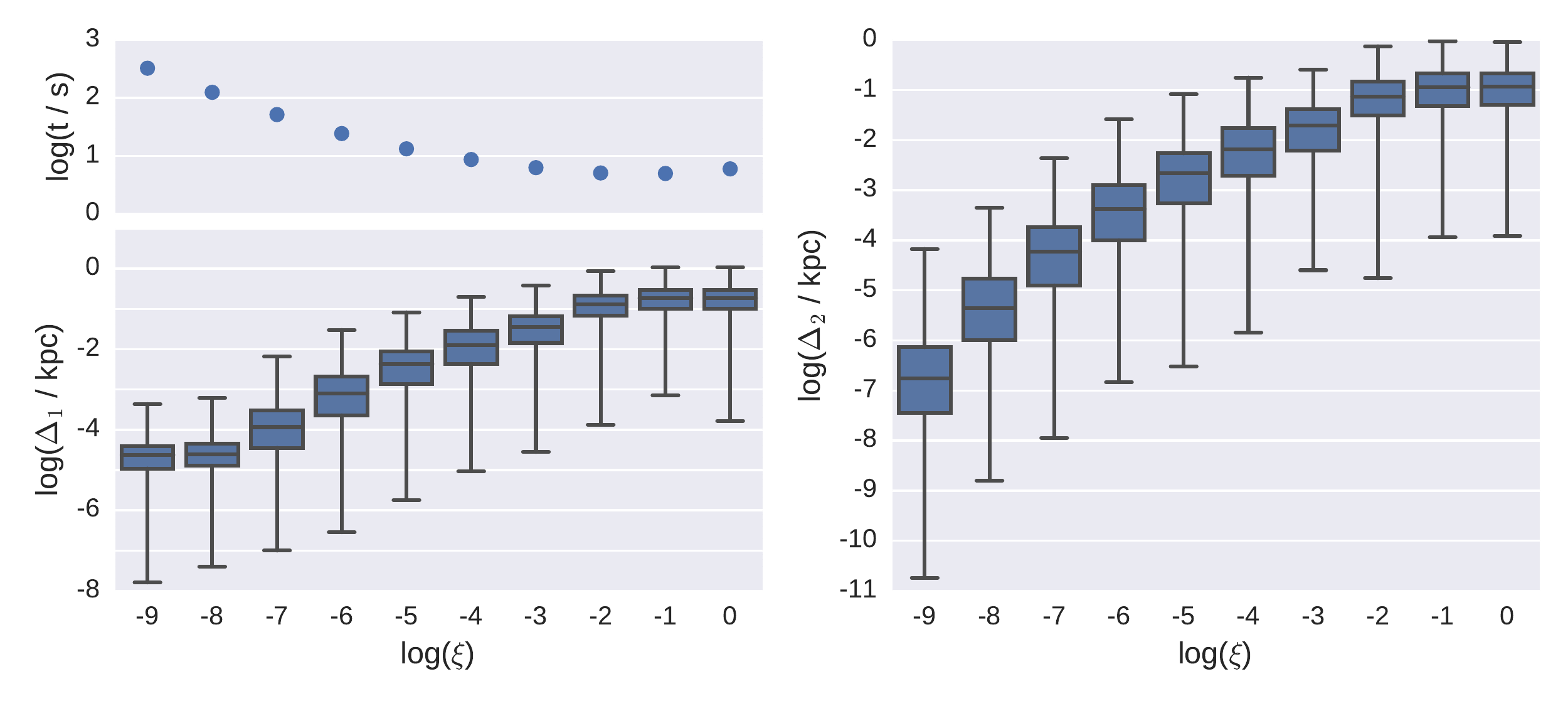}
 \caption{(Upper left) Computation time needed for 10{,}000 trajectories. (Lower left) Deviation 
from the expected position of the pseudo-particles per precision. (Right) 
Deviation from the spiral line. It is calculated as the minimum distance between particle 
position and the field line. The horizontal line marks the mean, the box indicates 
the range of fifty percent of the particles and the whiskers give the total range.}
 \label{fig:SpiralResults}
\end{figure}

In addition to the position of the 
pseudo-particles we also tracked the arc length $L_{sim}$ of the particles trajectories. The 
simulated arc length $L_{sim}$ is calculated as the sum of the propagation steps considering 
the sign and not just the absolute value. From this variable we are able to 
reconstruct the true particle position. We calculate the arc length 
$L$ for an the end position $z_{max}=z(T_{max})$ and fixed start position $z_0=0$:
\begin{align}
 L_{ana} &= \int_0^{z_{max}} \left|\left|\frac{{\rm d}r_{\rm spiral}(z)}{{\rm d}z}\right|\right| {\rm d}z \notag \\
 &= \int_0^{z_{max}}{\bigg(2 + \Big(\underbrace{ \frac{2\pi}{s}}_{=a} z \Big)^2\bigg)^{0.5}\,{\rm d}z} \notag \\
 &= \frac{1}{2}z_{\rm max}\sqrt{a^2 z_{\rm max}^2+2} + \frac{\sinh^{-1}\left(\frac{a z_{\rm max}}{\sqrt{2}}\right)}{a} \quad . \label{eq:ArcLength}
\end{align}
Since \Equ{eq:ArcLength} is not easily invertible we use a numerical solver to 
minimize $L_{sim}-L_{ana}(z)$. The best fitting $z_{ana}$  is used for 
the error estimation. Using $z_{ana}$ we can calculate the remaining two coordinates via 
\Equ{eq:Spiral}. This deviation from the analytic position $\Delta_1 = |\vec{r}_{sim}-\vec{r}_{ana}|$ is shown on the bottom left in \Fig{fig:SpiralResults}.

Furthermore, we calculate the absolute distance from the field line as an alternative measure of 
the algorithm accuracy. In doing so, we simply minimize the distance between the simulated 
particle end position $\vec{r}_{sim}$ and the field line. This leads to the second error $\Delta_2 
= \min(|\vec{r}_{sim}-\vec{r}_{spiral}|)$ which is shown on the right of \Fig{fig:SpiralResults}.

We repeated the test for precisions $\xi$ from $1$ to $10^{-9}$ and recorded the computation time for 
the total simulation\footnote{The simulation was done on a single core of the
Intel\textregistered Core i7-6800 chip at $3.4\,{\rm GHz}$.}. \Figure{fig:SpiralResults} shows the 
results of this test. In the upper left the computation time is shown and it is clearly visible 
that there is no linear correlation between the precision and the computation time. This is due to 
the adaptive algorithm which uses small integration steps at low $|z|$ and increases the 
integration time in regions where the curvature radius is larger. In contrast, the deviation from 
the field line is nearly linearly correlated to the precision in the region where it is not bound by 
the maximum integration step (precision of $0.01-1$). The deviation from position $\Delta_1$ saturates not 
only at the low precisions but also at very high precisions. This is most likely due to the 
restricted resolution of the numerical minimizer.
This test clearly shows that the new adaptive algorithm is able to follow a field line with 
changing curvature radius with a user defined precision. Since the curvature radii of the Galactic 
magnetic field are in most cases much larger than in this test a precision of order $10^{-4}$ is 
most likely sufficient to receive sub-parsec errors from the propagation.

In general, it should be noted that such a large range of possible integration steps is not the 
most 
effective way to use an adaptive algorithm. A very high accuracy of order $10^{-7}$ or 
higher will probably lead to many (up to the order of 20 or more) consecutive step refinements. 
This can lead to millions of field line integrations for a single diffusion step (see 
\Sec{ssec:technical} for details). The user should carefully choose a feasible range of steps and a 
reasonable precision. Good parameters are hard to predict but small benchmark simulations will 
likely help to find a useful set.

\subsubsection{Galactic Magnetic Field}
\label{sssec:JF12FieldlineIntegration}
From the tests described in \Sec{sssec:SpiralIntegration} it is known that the field line 
integration works. But what does that mean with respect to the Galactic environment? It is the aim 
of the proposed test here to evaluate whether the adaptive integration methods works on realistic 
Galactic field parameterizations.

The accuracy of the algorithm is supposed to be strongly correlated to the underlying magnetic field 
model. A simple continuously differentiable magnetic field is supposed to lead to a better result than 
a partly discontinuous field or one with poles. To get an estimate of the problems occurring from 
realistic Galactic magnetic fields, the magnetic field by Janson and Farrar (JF12-field) \cite{JAN12} was chosen for this test.

The magnetic field line is to be approximated with trajectory points along the magnetic 
field, again assuming $\kappa_\perp=0$. In doing so, magnetic field lines of $20\,{\rm kpc}$ length in $17\,{\rm pc}$ steps have been 
created for 100 randomly chosen emission points in the Galaxy.

In the following, a total number of $10^5$ particles are injected into the simulation. The start 
positions of the particles are randomly chosen among the 100 
different sources locations for which 'best fit` field lines exist. The end position of the 
particles are recorded. Now the deviation from the field line is approximated by the minimum of 
the differences of field line points and end positions. In this way, 100 sets of minimum distances 
are created with a sample size of around 1000 each. This procedure is done for five 
different rigidities $\rho=(10^{13}-10^{17})\,{\rm V}$.

To analyze the data the mean distance $<\Delta R>$ is calculated for each rigidity. 
\Table{tab:JF12Results} shows that the error increases slightly with increasing rigidity. Higher 
rigidities imply increased mean spatial steps for the same time interval. The adaptive algorithm 
normally accounts for this problem by a reduction of the integration time. When the minimum 
integration time is reached the error cannot be reduced to the optimum which is visible here for 
the highest rigidities. The error does not converge to zero due to the finite resolution of the 
ideal magnetic field lines which leads to expected minimal error of $\Delta R_{\rm ideal} = 
4.6-4.7\,{\rm pc}$.
\begin{table}[htbp]
\caption{Mean deviation from field line in the JF12 field for different rigidities}
\centering
\begin{tabular}{c|ccccc}
\toprule
$\rho\;[{\rm TV}]$& $10$ & $100$ & $1000$ & $10^4$ & $10^5$ \\ 
\midrule
$\langle\Delta R\rangle\; [{\rm pc}]$& 4.6 & 4.6 & 4.8 & 5.4 & 6.2 \\
\bottomrule
\end{tabular}
\label{tab:JF12Results}
\end{table}

Although the current implementation of the JF12 field is not completely continuously differentiable 
the adaptive algorithm works well. 

%% file: Example.tex
\section{Cosmic Ray Density Evolution}
\label{sec:Example}

There are many possible scenarios to use the new code, starting with the simulation of 
nearby known supernova remnants (SNRs) or a backtracking of particles, originating in the solar 
system to identify possible source regions inside the Galaxy. We decided to perform a different 
approach 
and simulate the diffusion of cosmic rays in a global sense. In order to get an estimate of the 
overall cosmic 
ray distribution we use a continuous source distribution following the SNR distribution given in 
\cite{Blasi2012I}. Furthermore, the realistic JF12 field, explained in 
\Sec{sssec:JF12FieldlineIntegration}, is 
used as the regular background field for the diffusion. Different values for the diffusion 
parameter $\epsilon$ with $\kappa_\perp=\epsilon\kappa_\parallel$ are tested, as well as the 
influence of the rigidity on the density evolution.

Since we fix the energy for each simulation we do not have to include any process that might change the energy of the pseudo-particles. For a realistic, full-scale simulation of the CRs in the Galaxy this should of course be included. These processes are for example energy losses by interaction but also diffusive re-acceleration, which could in principle be implemented as a diffusion process in momentum also using stochastic differential equations. 

Before we explain the results in \Sec{ssec:ExampleResults} the source distribution and the 
simulation setup are explained in  \Sec{ssec:SourceDistribution} and in 
\Sec{ssec:SimulationSetUp}, respectively.

\subsection{Source distribution}
\label{ssec:SourceDistribution}
We use the same supernova distribution as in \cite{Blasi2012I} where the radial distribution is 
originally given by Case and Bhattacharya \cite{Case1996}. Although this distribution just gives a 
rough estimate it
should reflect the general features of the SNR distribution inside our Galaxy well enough.
The radial distribution is described by:
\begin{align}
 f_R(r) = 
\frac{A_R}{R_0^2}\left(\frac{r}{R_0}\right)^2\exp\left(-\beta\frac{r-R_0}{R_0}\right) 
 \quad, \text{with}\quad 1 = \int_0^\infty 2\pi r\cdot f_R(r) \,{\rm d} r \label{eq:SNR_R}\quad ,
\end{align}
where $\beta=3.53$ and $R_0=8.5\,{\rm kpc}$ are taken from \cite{Blasi2012I} and 
$A_R=\beta^4\exp(-\beta)/(12\pi)$ is the normalization constant. The distribution in z-direction is 
defined by:
\begin{align}
 f_Z(z) = \frac{A_Z}{z_g}\exp\left(-\frac{|z|}{z_g}\right)\label{eq:SNR_Z} \quad ,
\end{align}
where $z_g=0.3\,{\rm kpc}$ is chosen as the scale height of the distribution and $A_Z=1$ is once 
more a normalization constant.

To generate the pseudo-particle start position we use a rejection sampling for $r$ and $z$ from 
\Equ{eq:SNR_R} and \Equ{eq:SNR_Z}, respectively. The third coordinate in cylinder 
coordinates, the angle $\phi$, is drawn from a uniform 
distribution with $\phi\in(0,2\pi]$. After that the coordinates are transformed into cartesian 
coordinates $x=r\cos(\phi)$, $y=r\sin(\phi)$ and $z=z$. Our approach is different from the 
one in \cite{Blasi2012I} since we do not define a given 
number of SNR positions drawn from the distribution but every particle has its own, unique origin. The 
sampling technique implemented for the source generation is very simple but fast enough. We can 
generate one million source positions in a few seconds, meaning that the generation of sources is 
not a significant part of the overall simulation time.

\begin{figure}
\centering
\begin{minipage}{0.5\textwidth}
\subfigure[][Radial distribution of SNR]{
\includegraphics[width = .9\textwidth]{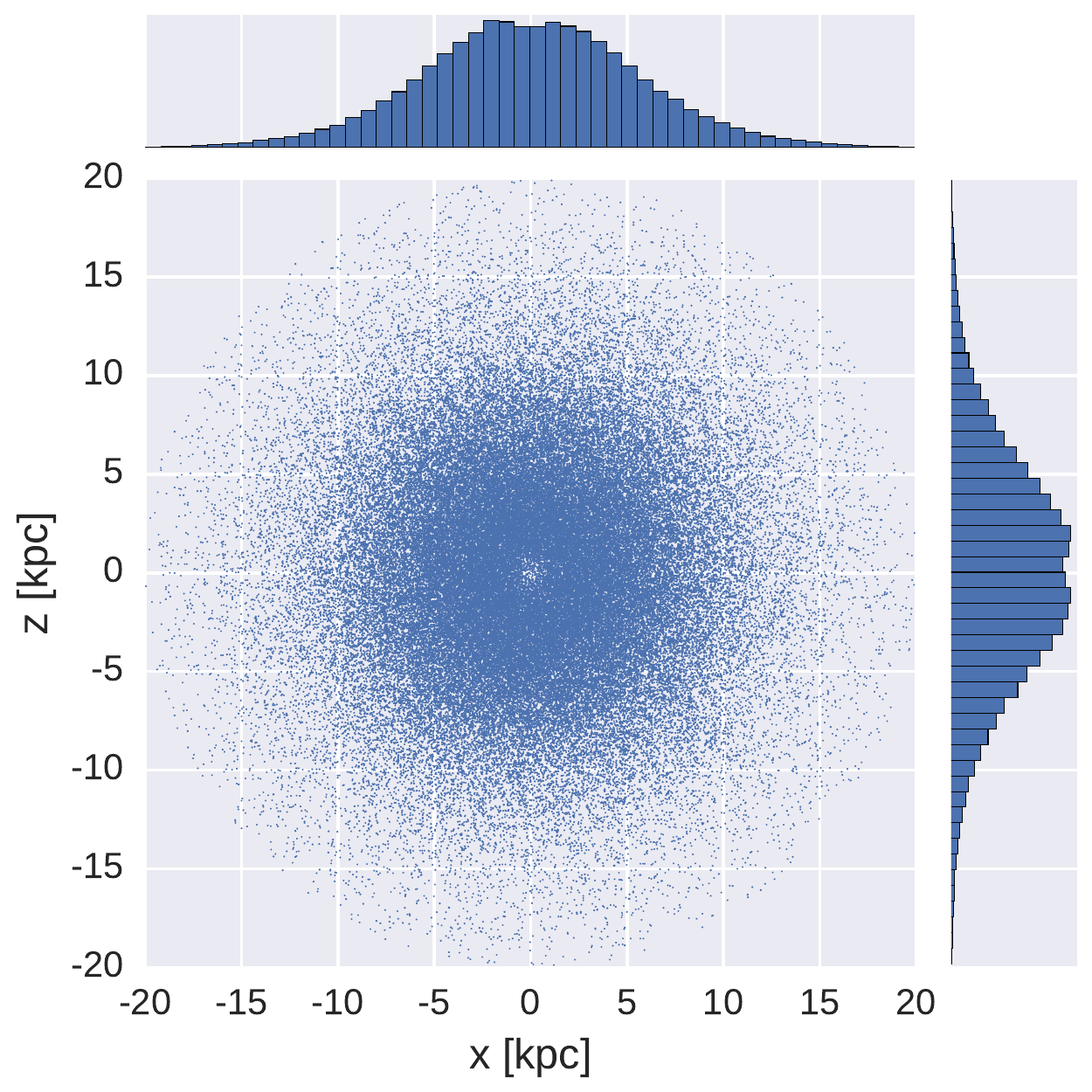}
\label{fig:SourceDistribution_FaceOn}
}
\end{minipage}%
\begin{minipage}{0.5\textwidth}
\subfigure[][Vertical distribution of SNR]{
\includegraphics[width = .9\textwidth]{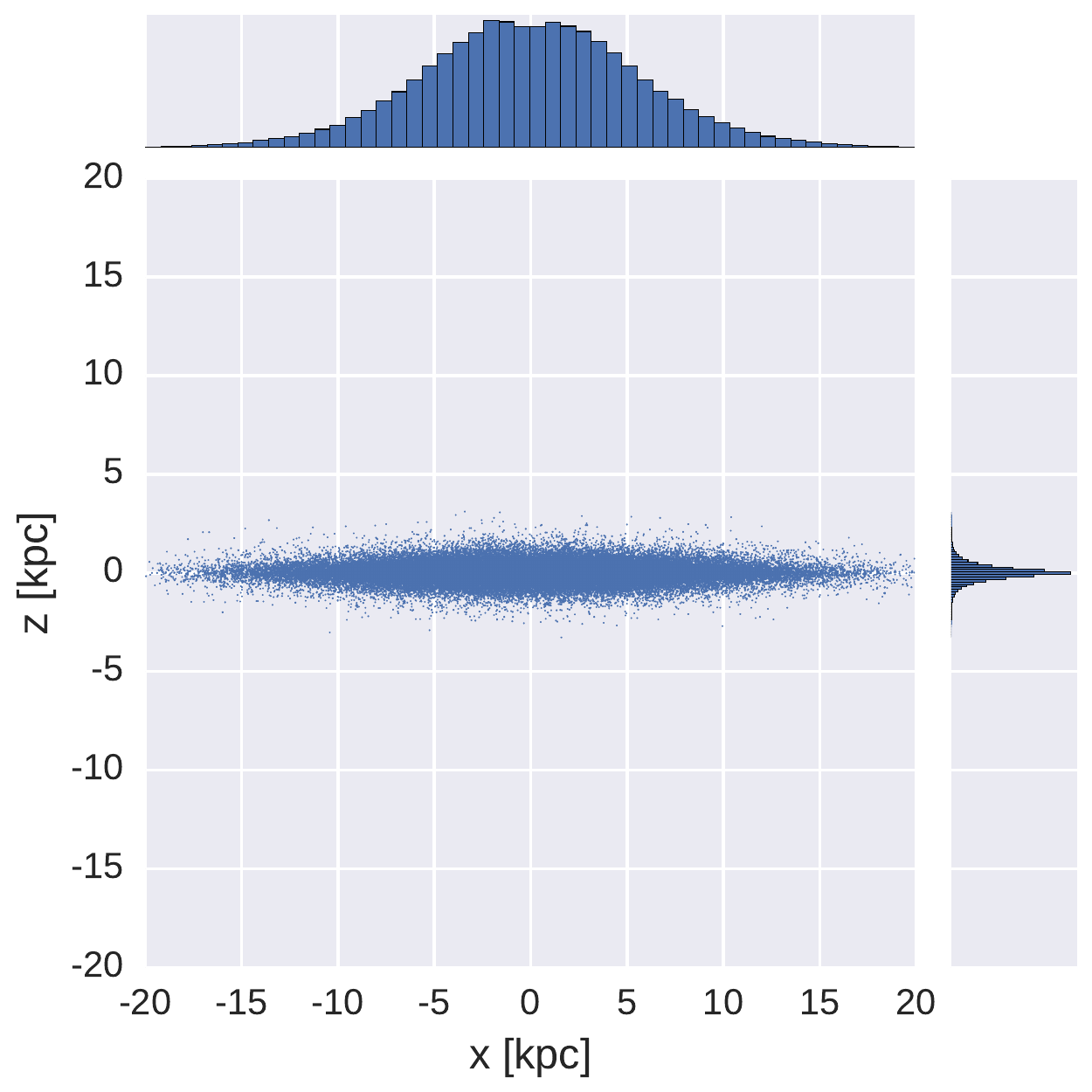}
\label{fig:SourceDistribution_EdgeOn}
}
\end{minipage}
\caption{The distribution of $10^6$ random source positions. On top and on the right side the 
corresponding one-dimensional histograms are displayed. There is no dependence of the scale height 
$z_g$ on $r$ but this is just a projection effect in figure (b).}
\label{fig:SourceDistribution}
\end{figure}

\Figure{fig:SourceDistribution} shows the distribution of SNRs used in the simulation. It is 
visible 
that the density has a minimum at $r_{min}=0$ and and maximum around 
$r_{max}=2R_0/\beta\approx4.8\,{\rm kpc}$. After that the density decreases exponentially and is 
cut off at the boundary of the simulation volume at $r>20\,{\rm kpc}$ (see 
\Sec{ssec:SimulationSetUp}). The rotation symmetry is clearly visible which should be kept in mind 
to interpret the results in \Sec{ssec:ExampleResults}.

\subsection{Simulation Setup}
\label{ssec:SimulationSetUp}
In this section the other simulation parameters, beside the source distribution, are discussed in 
detail. 

As mentioned above, the regular magnetic field component of the JF12 field 
\cite{JAN12} is used as the background field. Due to the fact that we only use the regular 
component we have to slightly adapt the field. In the original implementation of the field 
the regular field component of the field vanishes for $|\vec{r}| \leq 1\,{\rm kpc}$, where the origin 
$\vec{r}_0=0$ is at the Galactic center (meaning 
$B_{reg}=0$ in this region). Since this zero-field sphere has no physical reason but is due to a 
lack of precise observational data we continue the field to $\vec{r}_0=\vec{0}$. In doing so, we neglect the hard cut off at $|\vec{r}| = 1\,{\rm kpc}$ which is possible since the 
analytic 
form 
of the field is continuous up to $\vec{r}_0$. 

This simulation is primarily designed to study the influence of the diffusion parameter $\epsilon$ 
and rigidity $\rho$
on the cosmic ray distribution. The diffusion tensor $\hat{\kappa}$ is diagonal in the local frame 
of the 
magnetic field line, $\hat{\kappa} = {\rm diag}(\kappa_\perp, \kappa_\perp, \kappa_\parallel)$ with 
the parallel diffusion coefficient defined as in \cite{GALPROP}:
\begin{align}
 \kappa_\parallel = A \kappa_0 \left(\frac{\rho}{4 {\rm GV}}\right)^\alpha 
\label{eq:DiffusionCoefficient}\quad ,
\end{align}
where $\kappa_0=6.1\cdot10^{24}\,{\rm m^2/s}$, $\alpha=0.3$ and $A(\epsilon) = 1.02/(1+2\epsilon)$ 
is a normalization constant. The normalization $A$ is chosen to keep the trace of the diffusion 
tensor constant for different values of the diffusion parameter 
$\epsilon$. This is necessary since the value of the trace has a huge impact on the loss 
time scale. Using this normalization we exclude the influence of a varying trace on the time scale 
and are able to analyze the influence of the diffusion parameter $\epsilon$ or rigidity $\rho$.

In every simulation setup we calculate the trajectories for about 15 million pseudo particle  
forward in time until the maximum integration time $T_{max}=100\,{\rm Mpc/c}$ or the escaping from the 
simulation volume at $|\vec{r}_{max}|=20\,{\rm kpc}$. We use adaptive steps with 
$h_{min}=0.1\,{\rm pc/c}$, $h_{max}=1.0\,{\rm kpc/c}$ and precision $\xi=10^{-5}$. In that way we 
ensure the desired accuracy for the averaging process (see below). The 
particle position is recorded at 1000 points in time $t_n=n\Delta t$ with $\Delta t=100\,{\rm kpc/c}$ 
and $n=1...1000$. For the first analysis we use protons with a rigidity of $\rho=10\,{\rm TV}$ and for 
the second analysis we iteratively increase the proton rigidity by an order of magnitude up to 
$\rho=100\,{\rm 
PV}$. 

The simulation is performed on a cluster with 256 cores and takes about 3200 CPU hours per 
simulation set. The raw output files take a total memory storage of up to 300 GB per simulation 
set, depending on the escape time scale. To handle this huge amount of data we bin 
the pseudo-particle positions into a three-dimensional histogram with $1\,{\rm kpc}$ bin edges 
for each time 
step $t_n$ and simulation setup. This concurrently ensures that the particle density is 
averaged over a sufficiently large number of pseudo-particles.

\subsection{Results}
\label{ssec:ExampleResults}

The first results show that the diffusion along the magnetic field lines affect the cosmic 
ray density evolution significantly. We can show that even a nearly unstructured source 
distribution is transformed into a complicated density distribution following the magnetic field 
structure of the Galaxy. This is in contrast to earlier simulations where the source distribution 
itself is already shaped in a spiral structure e{.}g{.} \cite{GALPROP, Kissmann2014}. We discuss the 
results for varying diffusion parameter $\epsilon$ 
(\Sec{sssec:DiffusionParameter}) and rigidity $\rho$ (\Sec{sssec:Rigidity}) separately.

\subsubsection{Diffusion Parameter}
\label{sssec:DiffusionParameter}
\begin{figure}
\centering
 \includegraphics[width=\textwidth]{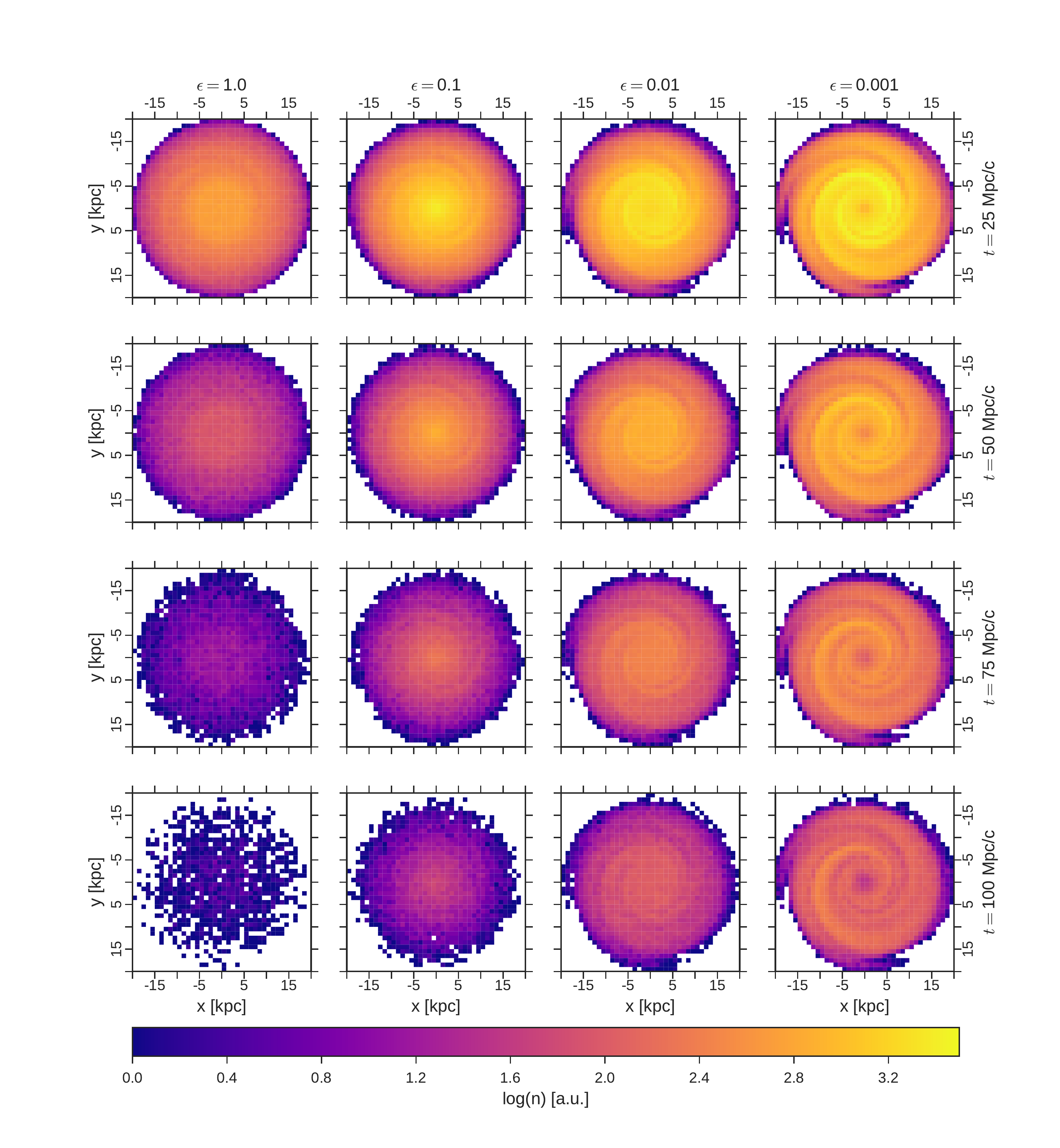}
 \caption{Face on view of the time evolution of the Galactic cosmic ray density. Here, only 
particles inside the Galactic disc ($-1\,{\rm kpc}\leq z\leq 1\,{\rm kpc}$) are displayed. From 
left to right the diffusion parameter $\epsilon$ is decreased. The time evolution is shown from 
top (early) to bottom (late). The density is given in arbitrary units on a log scale. Here, only protons with a rigidity of $E=10\,{\rm TV}$ are shown.}
 \label{fig:Density_xyProjection}
\end{figure}

\Figure{fig:Density_xyProjection} shows different face-on views (anti-parallel to the 
z-axis) of the Galactic disc. In order to highlight the Galactic structure we display only the 
particles near the Galactic plane with height $|z|\leq 1\,{\rm kpc}$. The difference in the 
diffusion parameter $\epsilon$ is clearly present. As expected, a decreased perpendicular diffusion 
coefficient emphasizes the structure of the magnetic field. In the case of very strong perpendicular 
diffusion $\kappa_\perp=\kappa_\parallel$ almost any structure, besides a fall off in the direction 
of the boundary, can be found in the density distribution. 

Another interesting fact is the 
difference in the cosmic ray density at the Galactic center. A strong perpendicular diffusion 
$\epsilon\geq 0.1$ leads to an increase of the Cosmic ray density. Finally, the 
maximum density region is shifted into the Galactic center. In contrast, strong parallel diffusion 
prevents the pseudo-particles from diffusion into the central region.

\begin{figure}
\centering
 \includegraphics[width=\textwidth]{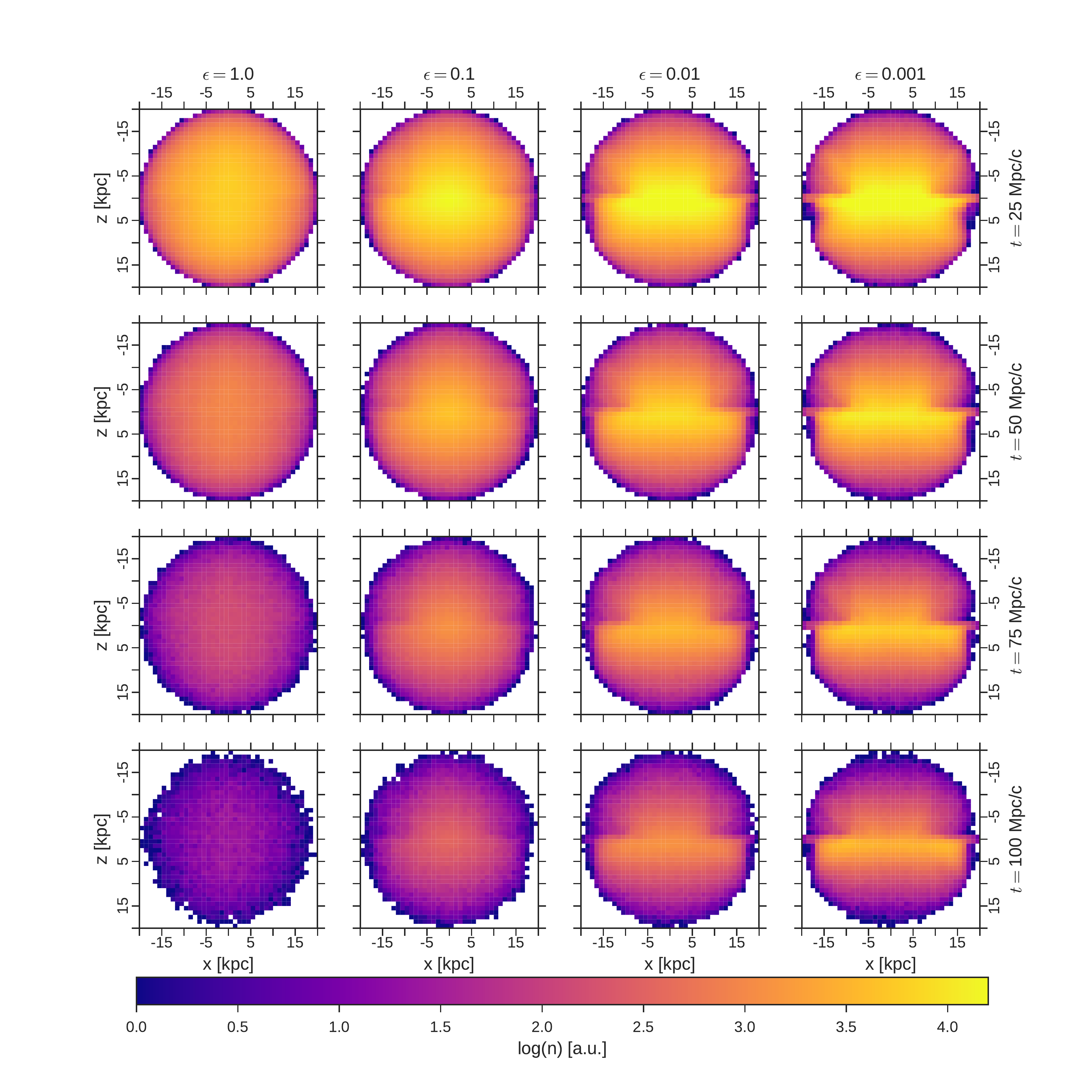}
 \caption{Edge on view of the time evolution of the Galactic cosmic ray density. Here, all 
particles inside the simulation volume are displayed. From 
left to right the diffusion parameter $\epsilon$ is decreased. The time evolution is shown from 
top (early) to bottom (late). The density is given in arbitrary units on a log scale. Here, only protons with a rigidity of $E=10\,{\rm TV}$ are shown.}
 \label{fig:Density_xzProjection}
\end{figure}

\begin{figure}
\centering
 \includegraphics[height=.3\textheight]{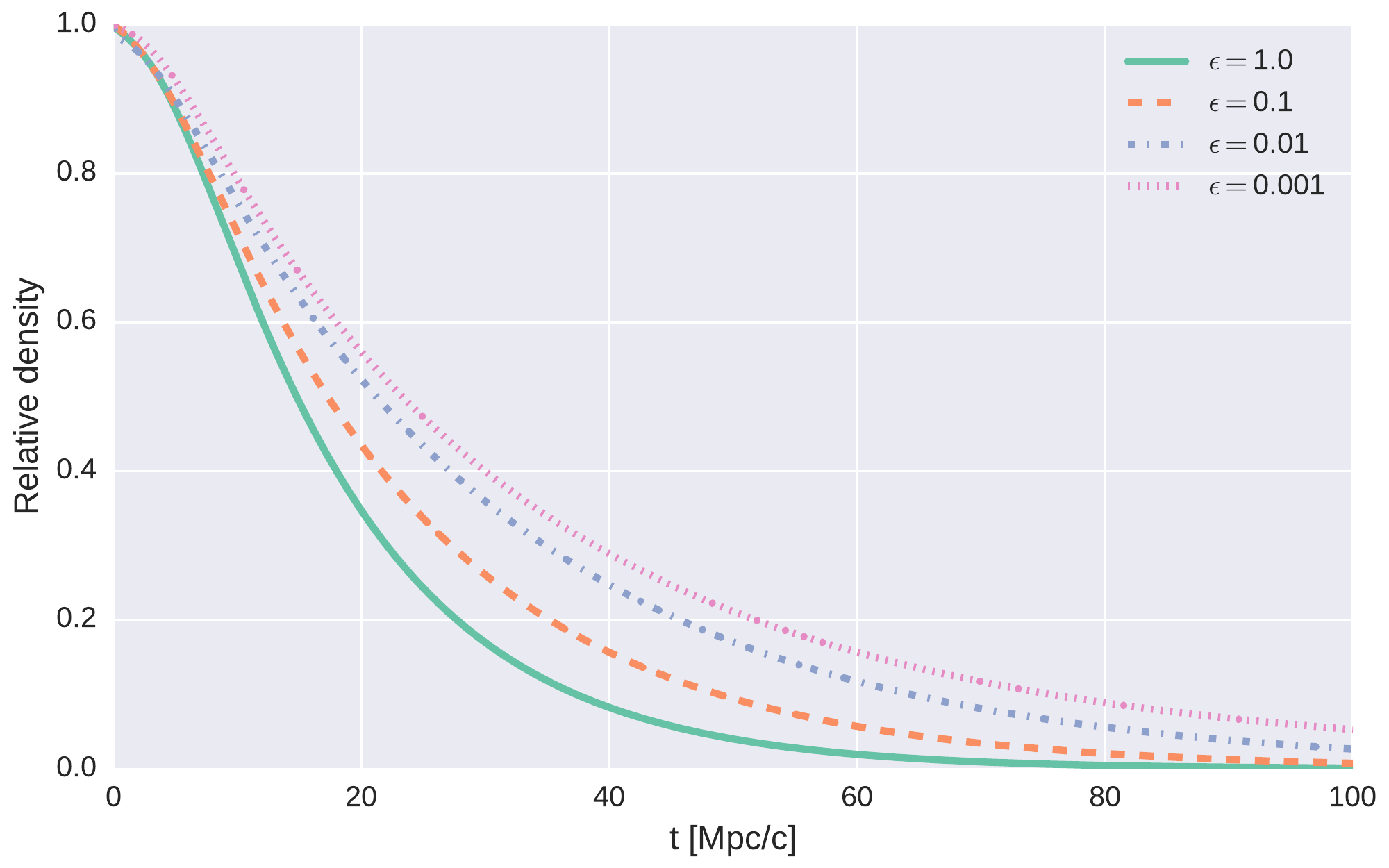}
 \caption{The total particle number relative to the number of injected particles is shown for the 
different simulation setups. Although the total diffusion strength ${\rm tr}(\hat{\kappa})$ is 
constant a significant difference in the escape time scale is visible.}
 \label{fig:Density_TimeEvolution}
\end{figure}

In addition to the analysis of the Galactic disc, we examined the cosmic ray density in the halo as 
well. In doing so, we combined in \Fig{fig:Density_xzProjection} all particles to produce 
edge-on representations of the Galaxy. As in the face-on projection the unstructured state of the
cosmic ray density for (very) strong perpendicular diffusion is clear. The general trend of higher 
escape rates in the case of increasing perpendicular diffusion is apparent and quantified in 
\Fig{fig:Density_TimeEvolution}. 

Furthermore, especially for strong parallel diffusion, a remarkable difference between the northern 
and southern hemisphere can be noticed. In the North the cosmic rays are concentrated in the center 
for lower Galactic latitudes. Perpendicular outflows are produced as clearly visible in the figures by the halo component of the magnetic field. These outflows have shapes that 
resemble the x-shape structure of the magnetic field that is detected for external galaxies (e.g.\ \cite{Heesen2009, Ferriere2014}) and their winds (e.g.\ \cite{Lopez2016}). In 
this paper, we do not perform a dedicated analysis in this direction and cannot draw any physics conclusions from these first findings. Detailed studies on the role of diffusion 
and advection for galactic outflows are planned in the future, but go beyond the scope of this paper. A similar structured outflow is 
not observed for the 
southern part of the Galaxy. This difference is an effect of the different magnetic field structure 
of the JF12 field in both hemispheres because the radial transition width of the toroidal field 
cannot be constrained by data for the southern hemisphere (see \cite{JAN12, FAR14} for details). 
This broken symmetry is not necessarily a real effect: the lack of observations for the halo field in the southern hemisphere leads to an approach to use the most simple magnetic 
field representation. Thus, even in the south, there might exist an x-shape-like structure, but it cannot be resolved by data yet at this point.

The escape time scale of the different diffusion configurations was mentioned above and is quantified 
in \Fig{fig:Density_TimeEvolution}. This is not surprising, although the different setups have 
equal overall diffusion strength, but describe a completely different morphology. This 
difference is mainly caused by the spiral magnetic field component which is strongly aligned in the 
Galactic plane. Since large perpendicular diffusion coefficients $\kappa_\perp \approx 
\kappa_\parallel$ allow a fast escape into the halo these configuration have shorter loss times. 
In contrast, a dominant parallel diffusion component $\kappa_\parallel\gg\kappa_\perp$ binds the 
pseudo-particles strongly inside the Galactic disc preventing a fast escape.
We define the effective escape time scale $T_{loss}$ by
\begin{align}
 \int_V n(T_{loss})\,{\rm d}V = \int_V n_0\cdot\exp(-1) \,{\rm d}V\quad ,
\end{align}
as the time scale on which the total particle density decreases by $1/e$. The 
different time scale are given in \Tab{tab:LossTimeScale}.
\begin{table}[htbp]
\caption{Effective escape time scale for different diffusion parameter $\epsilon$}
\centering
\begin{tabular}{c|cccc}
\toprule
$\epsilon$ & $1.0$ & $0.1$ & $0.01$ & $0.001$ \\
\midrule
$T_{loss}\;[{\rm Mpc/c}]$ & $19.2$ & $23.3$ & $29.4$ & $32.6$\\
\bottomrule
\end{tabular}
\label{tab:LossTimeScale}
\end{table}

\subsubsection{Rigidity dependence}
\label{sssec:Rigidity}
In this section the influence of the particle rigidity on the evolution of the cosmic ray density is 
discussed. We fix the diffusion parameter $\epsilon=0.01$ and vary the 
rigidity on a logarithmic scale from $\rho=10-10^5\,{\rm TV}$. The analysis is the same 
as described in \Sec{sssec:DiffusionParameter}. High rigidities lead to a very 
fast particle loss, so we decided to show the time evolution on a logarithmic scale rather than on a 
linear scale (as in 
\Sec{sssec:DiffusionParameter}). 

\begin{figure}
\centering
 \includegraphics[width=\textwidth]{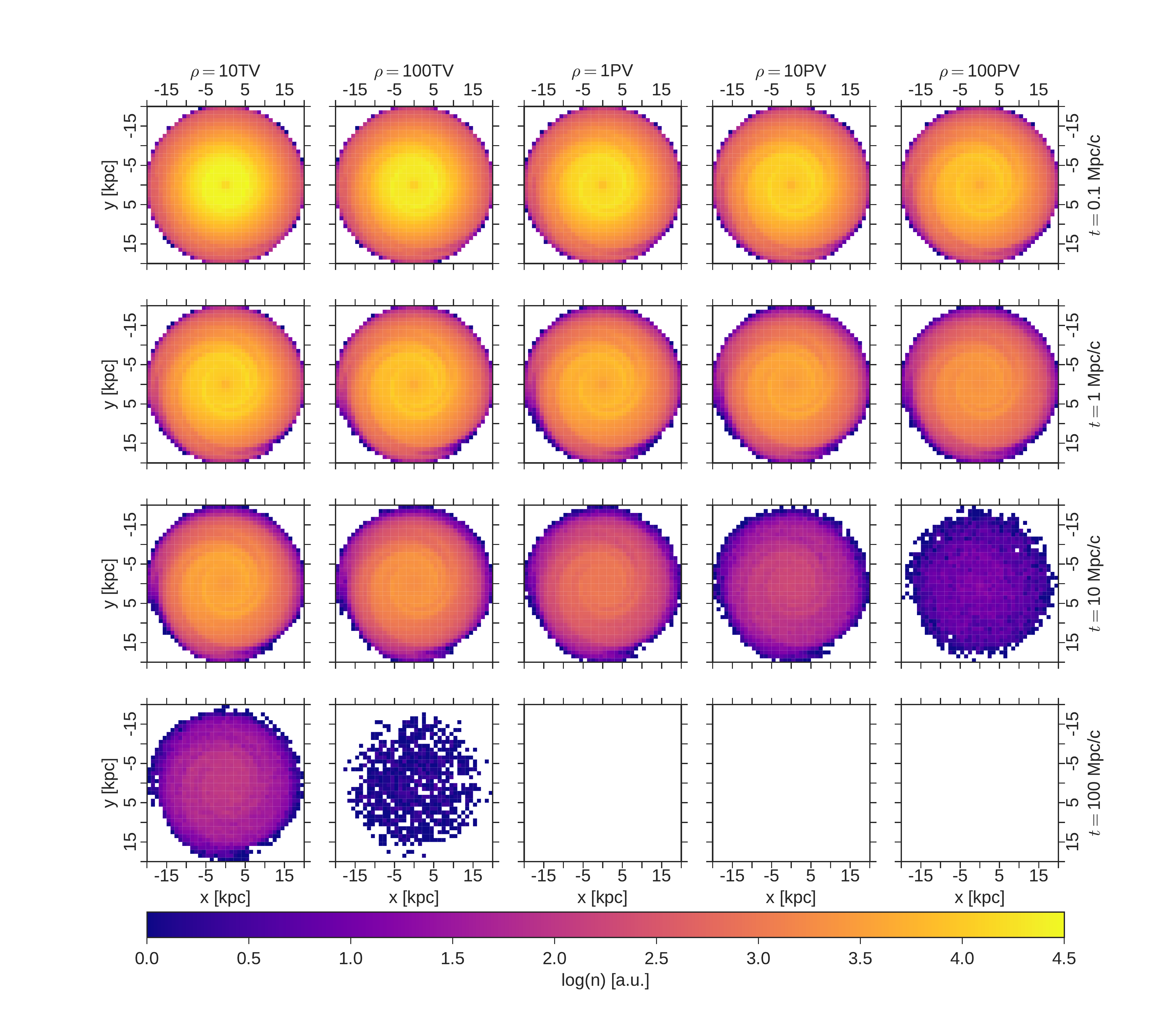}
 \caption{Face on view of the time evolution of the Galactic cosmic ray density. Here, only 
particles inside the Galactic disc ($-1\,{\rm kpc}\leq z\leq 1\,{\rm kpc}$) are displayed. From 
left to right the rigidity $\rho$ is increased. The time evolution is shown from 
top (early) to bottom (late). Here, the diffusion parameter $\epsilon=0.01$ is fixed. The density is given in arbitrary units on a log scale. Please notice the different points in 
time and density scales compared to 
\Fig{fig:Density_xyProjection}}
 \label{fig:Density_xyProjection_energy}
\end{figure}

\Figure{fig:Density_xyProjection_energy} shows the density distribution in the Galactic disc at 
$t=(0.1, 1, 10, 100)\,{\rm Mpc/c}$ for different rigidities. The very left column 
may help to compare the two analyses as the data for $\rho=10\,{\rm TV}$ was already used in the 
previous analysis. It is visible that the simulation for different rigidities do not differ much from 
each other when 
the general topology of the density distribution is compared. But it is obvious that an increase in 
the particle rigidity leads to an acceleration of the escape process. The 
different simulations are very similar to each other but on very different time scales.

\begin{figure}
\centering
 \includegraphics[width=\textwidth]{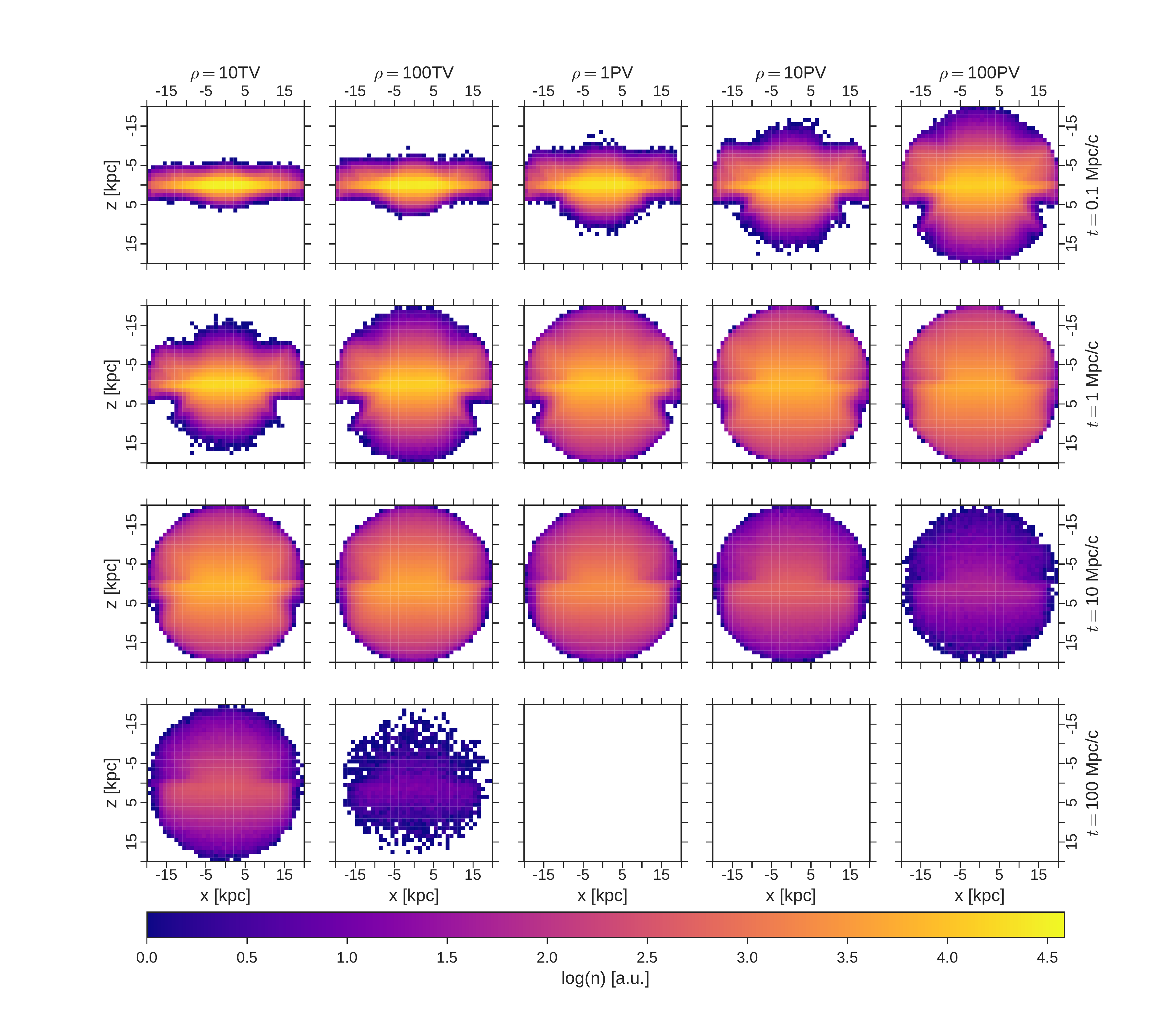}
 \caption{Edge on view of the time evolution of the Galactic cosmic ray density. Here, all 
particles inside the simulation volume are displayed. From 
left to right the rigidity $\rho$ is increased. The time evolution is shown from 
top (early) to bottom (late). Here, the diffusion parameter $\epsilon=0.01$ is fixed. The density is given in arbitrary units on a log scale. Please notice the different points in 
time and density scales compared to 
\Fig{fig:Density_xzProjection}}
 \label{fig:Density_xzProjection_energy}
\end{figure}

The effect of different time scales becomes even more prominent in the edge-on view shown in 
\Fig{fig:Density_xzProjection_energy}. At the highest rigidity the particles fill up
nearly the whole halo in just a few hundred thousand years. The early time evolution reveals a 
feature of the cosmic ray transport that was not visible in the first analysis. The 
x-shape like structure which is observed for the northern hemisphere at later time is also visible in the southern hemisphere but on much shorter time scale. After a couple 
million years, depending on the rigidity, the diffusion in the toroidal magnetic field washes the x-shape structures out.

\begin{figure}
\centering
 \includegraphics[height=.3\textheight]{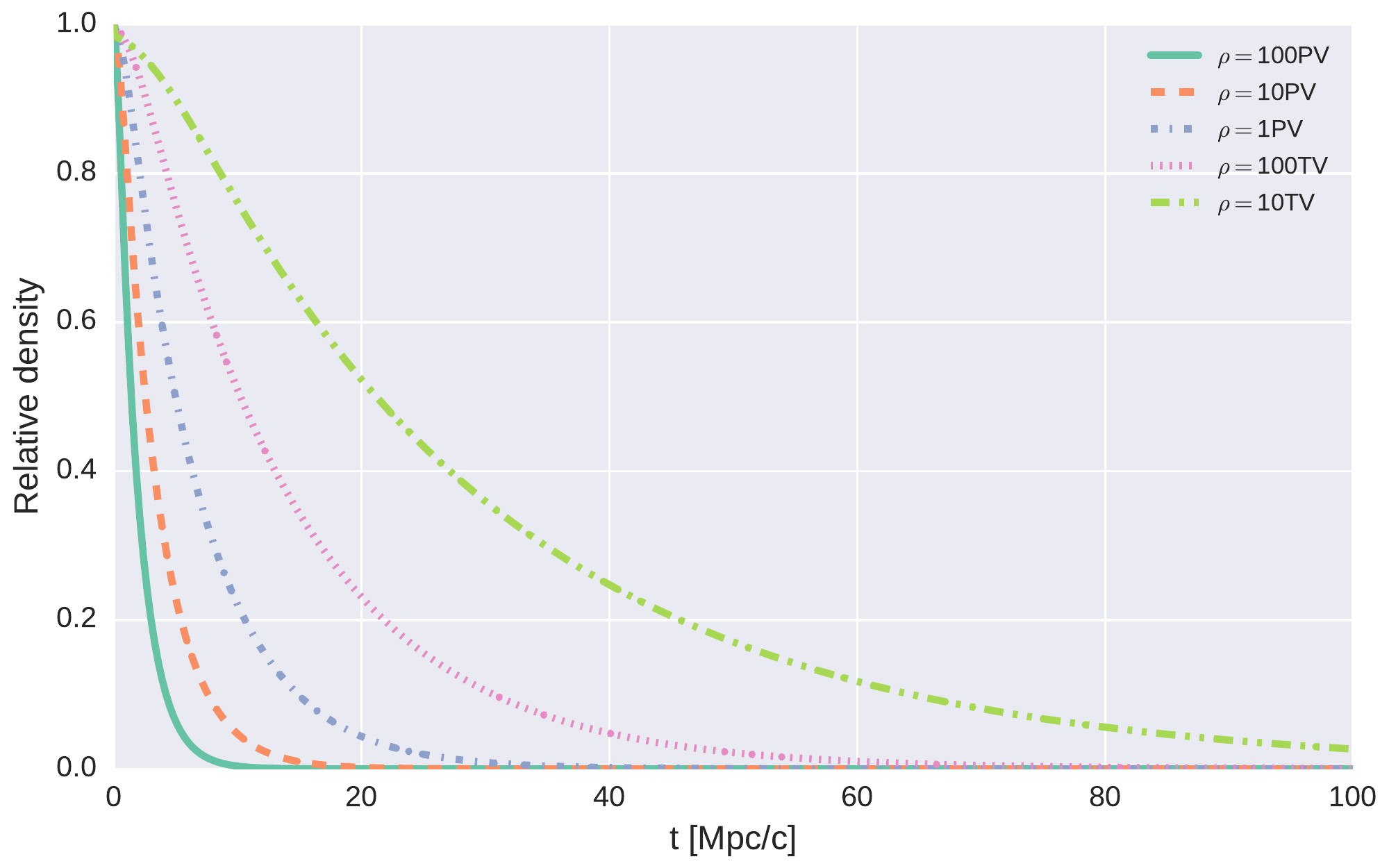}
 \caption{The total particle number relative to the number of injected particles is shown for the 
different simulation setups. As expected the escape time scale is strongly correlated to the rigidity of the particles.}
 \label{fig:Density_TimeEvolution_energy}
\end{figure}

\Figure{fig:Density_TimeEvolution_energy} emphasizes the huge differences in the escape time scales. for rigidities above $\rho=1\,{\rm PV}$ it takes less than a hundred million 
years to lose nearly all particles to the extra-galactic medium. 
In \Tab{tab:LossTimeScale_energy} the escape time scales are listed for a quantified analysis. 
Comparing the time scales we obtain that
\begin{align}
 T_{loss} \propto \rho^{-0.29} \quad ,
\end{align}
which reflects the rigidity dependence of the diffusion coefficient given in 
\Equ{eq:DiffusionCoefficient}.
\begin{table}[htbp]
\caption{Effective escape time scale for different rigidities $\rho$ and $\epsilon=0.01$}
\centering
\begin{tabular}{c|ccccc}
\toprule
$\rho\;[{\rm TV}]$& $10$ & $100$ & $1000$ & $10^4$ & $10^5$ \\ 
\midrule
$T_{loss}\;[{\rm Mpc/c}]$ & $29.4$ & $14.1$ & $6.9$ & $3.6$ & $2.0$\\
\bottomrule
\end{tabular}
\label{tab:LossTimeScale_energy}
\end{table}

%% file: Summary.tex
\section{Summary and Outlook}
\label{sec:SummaryOutlook}

In this section we emphasize once more the advantages of the new algorithm and recapitulate what is 
already 
possible using SDEs for Galactic propagation. In addition, we give a short outlook into the further 
development of simulation tasks and possible improvements of the status quo. 

In principle the diffusion approach as it is implemented here is also able to simulate CRs with energies below $E\leq10\,{\mathrm TeV}$. The problem here is that CRPropa uses approximations for very high energies when it comes to interactions processes. For example secondary particles are ejected parallel to the parent nucleus. This makes a low energy approach not impossible but surely the interaction modules have to revised.

\subsection{Summary}
\label{ssec:Summary}

In this paper we show that stochastic differential equations provide an efficient mathematical tool to describe the cosmic ray transport in the diffusive regime. This method is not 
a completely 
new approach but is already established in a few propagation codes used for the simulation in the heliosphere (e.g.\ \cite{Effenberger2012, Kopp2012}). For 
Galactic propagation only first examples exist (e.g.\ \cite{Effenberger2012a, Miyake2015}). But in contrast to the approaches mentioned above we do 
not solve the transport equation in the global or laboratory frame. As explained in detail in 
\Sec{ssec:Code} we take advantage of the fact that the diffusion tensor is diagonal in the frame of 
the coherent magnetic background field. This allows us to decouple the diffusion process for the three 
distinct directions of the orthormal basis defined by the comoving trihedron $\{\vec{e}_t, 
\vec{e}_n, \vec{e}_b \}$ (see \Equ{eq:trihedron}). In doing so, the diffusion 
tensor is completely defined by two parameters, the parallel diffusion coefficient 
$\kappa_\parallel$ and the diffusion ratio $\epsilon$.

We prove that the adaptive field line integration works in simple fields 
(\Sec{sssec:SpiralIntegration}) as well as realistic field descriptions 
(\Sec{sssec:SpiralIntegration}). Furthermore, a method to yield stationary solutions of the 
transport equation by averaging the time-dependent solutions is given and explained in 
\Sec{sssec:PICARD}. 

In \Sec{sec:Example} first basic simulation examples show the potential of this new simulation 
software. Already this very simple source distribution leads to interesting insights of the time 
evolution of the cosmic ray density in the Galactic halo. These need to be quantified and analyzed in future work.

We conclude that this software may help to develop a code to describe the transition region 
consistent in a single propagation framework. In this way systematic differences between the 
treatment of Galactic and extra-galactic cosmic rays can be minimized.

Finally, we want to emphasize that this propagation tool is restricted to Galactic propagation but 
can be used in any environment where the diffusion approximation holds. In principle, this 
technique may also be used for the simulation of reacceleration by the implementation of momentum 
diffusion.

\subsection{Outlook}
\label{ssec:Outlook}

In this section we will give a brief outlook on the future development and possible applications of 
the described tool. 

In the case of our own Galaxy we will examine the cosmic ray density 
evolution further. In doing so, we will use different source scenarios, e.g.\ applying discrete SNRs and not a 
continuous source distribution or more sophisticated source models taking the spiral 
structure into account (e.g.\ the distribution of isolated radio pulsars \cite{Faucher2006}).

In the future this new program will contribute to the solution of the gradient problem of the observed gamma-rays (e.g.\ \cite{Ackermann2011}) with the possibility of combining 
dedicated propagation tools 
with precise interaction models. The density maps shown in \Sec{sec:Example} are first step into the modeling of the cosmic ray gradients. The implemented anisotropic diffusion 
tensor will also help to understand the anisotropy of cosmic rays better.

From the technical point of view, several major extensions of the 
existing software are planned.
The new propagation software has some advantages over the grid-based methods as mentioned above. 
But up to now the software is not able to take all kinds of interactions, like proton-proton or proton-nucleon inelastic scattering processes into account. Only proton-photon 
interaction are already implemented very efficiently in the framework. This means in effect, a complete 
description of the 
transport of cosmic rays in a dense environment like our Galaxy is not yet possible. The implementation of the missing interaction processes is one 
of the next steps to improve the code. In doing so, we will adapt the methods already used and 
tested in CRPropa for proton-photon scattering. Here, we will use Monte-Carlo methods to generate 
secondary particles depending on the specific cross sections (see e.g.\ \cite{Mazziotta2016, Webber2003}) and 
inject them as new candidates 
into the simulation chain. Furthermore, density maps of the Galactic mass distribution have to 
implemented for a proper, location-dependent simulation of the spallation processes. Beside the implementation of the scattering processes three dimensional density maps of the 
target material (e.g.\ HI and HII maps from \cite{Nakanishi2016}) are crucial to derive realistic 
maps of neutral secondaries, such as gamma rays and neutrinos. We want to emphasise that 
implementation of interactions is relevant to model the fluxes of secondary particles like lithium, 
beryllium, boron, nitrogen, electrons, positrons
and neutrinos. On the other hand the flux of primaries, such as protons, will not be affected much 
by interactions because the interaction probability is much smaller than unity for the majority of 
the CR population. An exception would be the combined case of strong parallel diffusion 
($\epsilon\leq0.01$) and low rigidities ($\rho\leq100\,{\rm TV}$) where
primary fluxes can be significantly influenced by interactions.

To take Galactic winds or in general advection processes into account an additional term of the 
transport equation has to be factored in. The advection is described by the linear term $A$ in 
\Equ{eq:SDE} which means that the Euler-Maruyama Scheme in \Equ{eq:EMScheme} is modified by a 
further term proportional to the integration time step $h$.

Up to now the diffusion tensor is locally independent in the frame of the magnetic field line. This 
means that the diffusion tensor, transformed in the lab frame, has constant eigenvalues. Recent 
studies by Snodin and others have shown that the calculation of the diffusion tensor depending on the local magnetic field structure and the particle's gyro radius is possible 
\cite{Snodin2016}. They used test particle simulation to derive the diffusion tensor for a homogeneous background field and different models for the random magnetic field 
component. They give analytic formulas for the parallel and the perpendicular diffusion coefficient which might serve as a good starting point for a spatially varying 
implementation of the diffusion tensor in CRPropa 3.1. 

The technically most challenging problem we try to solve is the development of a dynamic switching between 
the two different propagation models. As mentioned in \Sec{ssec:ProgramStructure} the definition of 
parameters for the change between the propagation modes is non trivial. We will use machine learning 
algorithms to find the best simulation procedure depending on the current candidate properties, like particle type, position and rigidity. First 
studies may for example compare the escape times in equal environments using the different modules. 
The long term goal is to develop a software which is able to dynamically choose the best 
propagation module depending on user set criteria. These criteria should take limited computing 
resources as well as the specific physical simulation into account.

%% file: Appendix.tex
\section{Spatial error resolution stationary solution}
\label{ap:SpatialError}

\begin{figure}[htb]
\centering
 \includegraphics[height=.8\textheight]{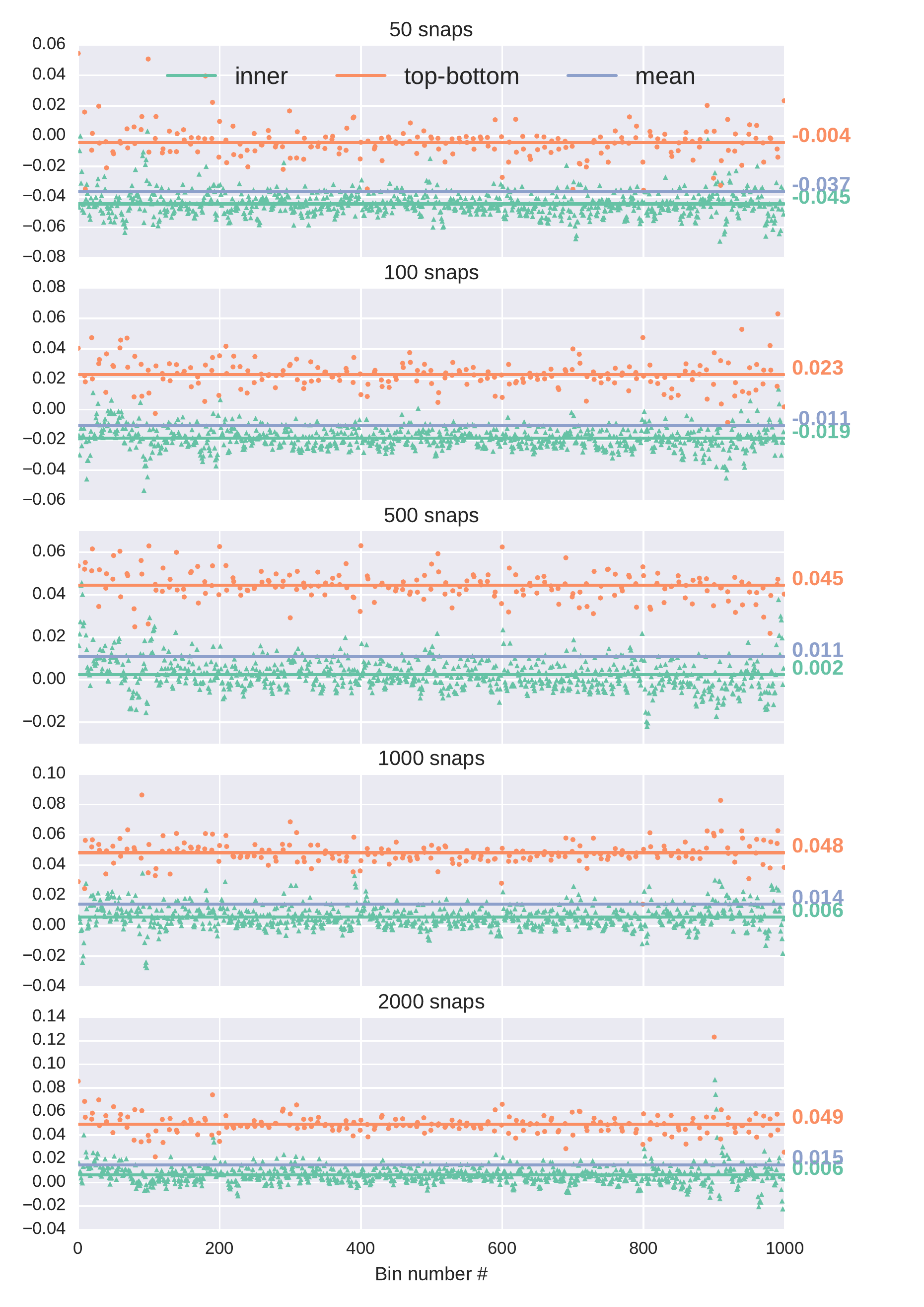}
 \caption{Relative error of the simulated density for different number of snapshots $N_{snap}$. The 
deviations are split into groups of bins at the top and bottom of the simulation box (orange, 
circle) and the inner part of the volume (green, triangles) (see \Fig{fig:Binning}). Orange and green lines indicate 
the mean of the groups and the blue one the overall mean.}
 \label{fig:PICARD_perBin_fixedStep}
\end{figure}

\begin{figure}[htb]
\centering
 \includegraphics[height=.8\textheight]{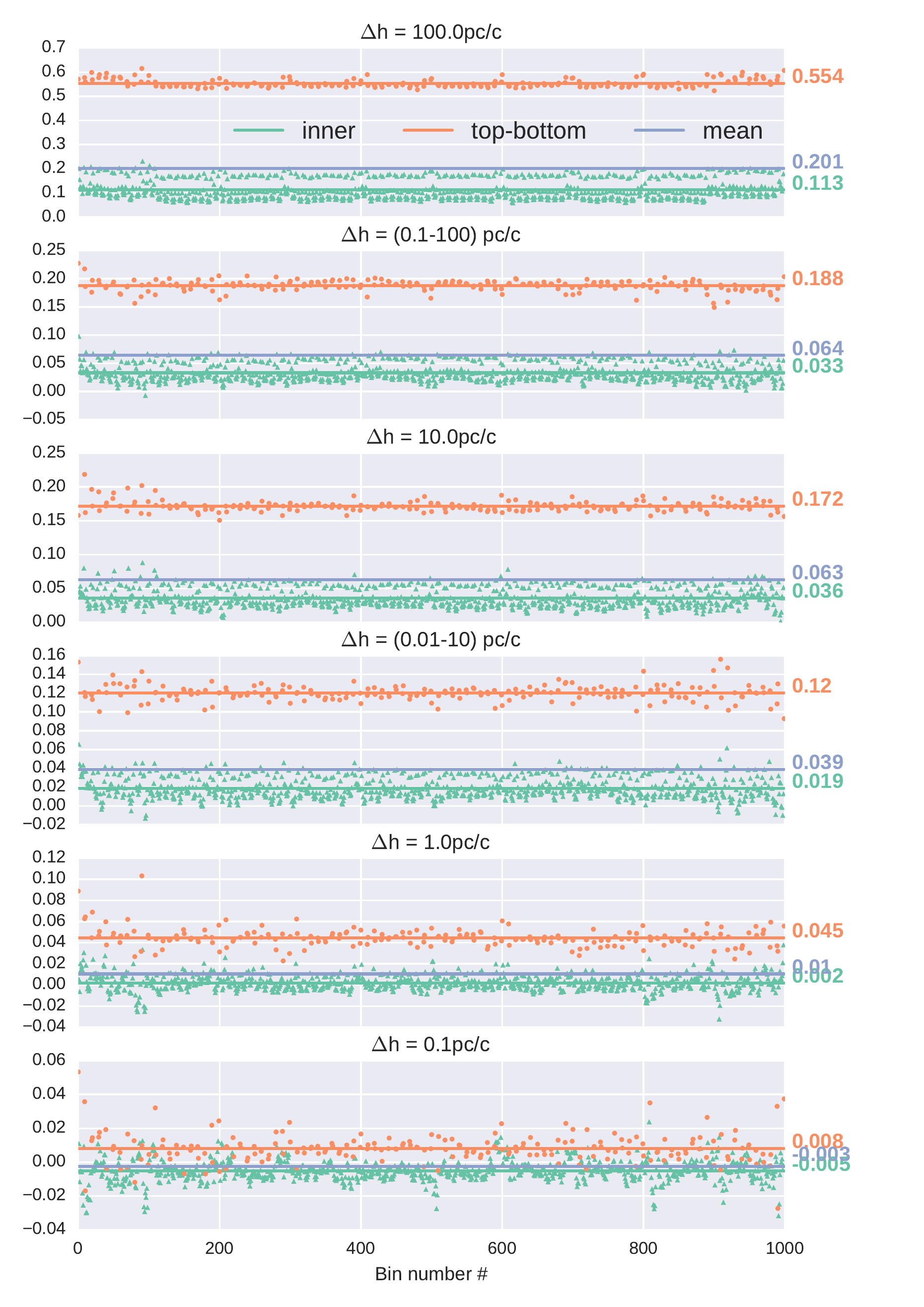}
 \caption{Relative error of the simulated density for different integration time steps $\Delta h$. 
The deviations are split into groups of bins at the top and bottom of the simulation box (orange, 
circle) and the inner part of the volume (green, triangles) (see \Fig{fig:Binning}). Orange and green lines indicate 
the mean of the groups and the blue one the overall mean.}
 \label{fig:PICARD_perBin_fixedSnap}
\end{figure}

%% file: Main.bbl
\providecommand{\href}[2]{#2}\begingroup\raggedright\begin{thebibliography}{10}

\bibitem{icecube_concept}
A.~{Achterberg} et~al., \emph{{First year performance of the IceCube neutrino
  telescope}},
  \href{http://dx.doi.org/10.1016/j.astropartphys.2006.06.007}{\emph{Astropart.~Phys.}
  {\bf 26} (Oct., 2006) }.

\bibitem{icecube2014}
M.~G. {Aartsen} et~al., \emph{{Observation of High-Energy Astrophysical
  Neutrinos in Three Years of IceCube Data}},
  \href{http://dx.doi.org/10.1103/PhysRevLett.113.101101}{\emph{Phys.~Rev.~Lett.}
  {\bf 113} (Sept., 2014) }.

\bibitem{hess_telescopes2003}
K.~{Bernl{\"o}hr} et~al., \emph{{The optical system of the H.E.S.S. imaging
  atmospheric Cherenkov telescopes. Part I: layout and components of the
  system}},
  \href{http://dx.doi.org/10.1016/S0927-6505(03)00171-3}{\emph{Astropart.~Phys.}
  {\bf 20} (Nov., 2003) }.

\bibitem{magic2004}
C.~{Baixeras} et~al., \emph{{Commissioning and first tests of the MAGIC
  telescope}}, \href{http://dx.doi.org/10.1016/j.nima.2003.10.057}{\emph{Nucl.\
  Inst.\ and Meth.\ in Phys.\ Res.\ A} {\bf 518} (Feb., 2004) }.

\bibitem{veritas2004}
F.~{Krennrich} et~al., \emph{{VERITAS: the Very Energetic Radiation Imaging
  Telescope Array System}},
  \href{http://dx.doi.org/10.1016/j.newar.2003.12.050}{\emph{New Astron.~Rev.}
  {\bf 48} (Apr., 2004) }.

\bibitem{Abraham2010}
J.~Abraham et~al., \emph{{Measurement of the energy spectrum of cosmic rays
  above $10^{18}$ eV using the Pierre Auger Observatory}},
  \href{http://dx.doi.org/10.1016/j.physletb.2010.02.013}{\emph{Physics
  Letters, Section B: Nuclear, Elementary Particle and High-Energy Physics}
  {\bf 685} (Mar., 2010) }.

\bibitem{ta_spectrum2015}
T.~{Abu-Zayyad} et~al., \emph{{Energy spectrum of ultra-high energy cosmic rays
  observed with the Telescope Array using a hybrid technique}},
  \href{http://dx.doi.org/10.1016/j.astropartphys.2014.05.002}{\emph{Astropart.~Phys.}
  {\bf 61} (Feb., 2015) }.

\bibitem{ta_composition2015}
R.~U. {Abbasi} et~al., \emph{{Study of Ultra-High Energy Cosmic Ray composition
  using Telescope Array's Middle Drum detector and surface array in hybrid
  mode}},
  \href{http://dx.doi.org/10.1016/j.astropartphys.2014.11.004}{\emph{Astropart.~Phys.}
  {\bf 64} (Apr., 2015) }.

\bibitem{Weng2016}
Z.~Weng and V.~Vagelli, \emph{Ams-02 measurement of cosmic ray positrons and
  electrons},
  \href{http://dx.doi.org/10.1016/j.nuclphysbps.2015.09.068}{\emph{Nuclear and
  Particle Physics Proceedings} {\bf 273} (Apr., 2016) }.

\bibitem{GALPROP}
A.~Strong and I.~Moskalenko, \emph{{Propagation of Cosmic-Ray Nucleons in the
  Galaxy}}, \href{http://dx.doi.org/10.1086/306470}{\emph{Astroph.~J.} {\bf
  509} (Dec., 1998) }.

\bibitem{Evoli2008}
C.~Evoli, D.~Gaggero, D.~Grasso and L.~Maccione, \emph{{Cosmic ray nuclei,
  antiprotons and gamma rays in the galaxy: a new diffusion model}},
  \href{http://dx.doi.org/10.1088/1475-7516/2008/10/018}{\emph{J.~Cosm.~and~Astr.~Phys.}
  {\bf 10} (Oct., 2008) }.

\bibitem{Kissmann2014}
R.~Kissmann, \emph{{PICARD: A novel code for the Galactic Cosmic Ray
  propagation problem}},
  \href{http://dx.doi.org/10.1016/j.astropartphys.2014.02.002}{\emph{Astropart.~Phys.}
  {\bf 55} (Mar., 2014) }.

\bibitem{V.S.1990}
V.~L. Ginzburg, V.~A. Dogiel, V.~S. Berezinsky, S.~V. Bulanov and V.~S.
  Ptuskin, \emph{{Astrophysics of cosmic rays}}.
\newblock North--Holland, 1990.

\bibitem{eichmann2016}
B.~{Eichmann} and J.~{Becker Tjus}, \emph{{The Radio-Gamma Correlation in
  Starburst Galaxies}},
  \href{http://dx.doi.org/10.3847/0004-637X/821/2/87}{\emph{Astroph.~J.} {\bf
  821} (Apr., 2016) }.

\bibitem{yoast_hull2013}
T.~M. {Yoast-Hull} et~al., \emph{Winds, clumps, and interacting cosmic rays in
  m82}, \href{http://dx.doi.org/10.1088/0004-637X/768/1/53}{\emph{Astroph.~J.}
  {\bf 768} (May, 2013) }.

\bibitem{schlickeiser_jenko2010}
R.~{Schlickeiser} and F.~{Jenko}, \emph{{Cosmic ray transport in non-uniform
  magnetic fields: consequences of gradient and curvature drifts}},
  \href{http://dx.doi.org/10.1017/S0022377809990444}{\emph{Journal of Plasma
  Physics} {\bf 76} (Aug., 2010) }.

\bibitem{Shalchi2009}
A.~{Shalchi}, \emph{{Nonlinear Cosmic Ray Diffusion Theories}}, vol.~362 of
  \emph{Astrophysics and Space Science Library}.
\newblock Springer, 2009.

\bibitem{Busching2011}
I.~B\"usching, A.~Kopp, F.~Effenberger, R.~Strauss, M.~Potgieter and
  H.~Fichtner, \emph{{A stochastic approach to galactic propagation}},  in
  \emph{International Cosmic Ray Conference}, vol.~6, (Beijing), 2011.
\newblock \href{http://dx.doi.org/10.7529/ICRC2011/V06/0841}{DOI}.

\bibitem{Effenberger2012}
F.~Effenberger, H.~Fichtner, K.~Scherer, S.~Barra, J.~Kleimann and R.~D.
  Strauss, \emph{A generalized diffusion tensor for fully anisotropic diffusion
  of energetic particles in the heliospheric magnetic field},
  \href{http://dx.doi.org/10.1088/0004-637X/750/2/108}{\emph{Astroph.~J.} {\bf
  750} (Apr., 2012) }.

\bibitem{gardiner2009}
C.~Gardiner, \emph{Stochastic Methods: A Handbook for the Natural and Social
  Sciences}.
\newblock {Springer Series in Synergetics}. Springer Berlin Heidelberg, 2009.

\bibitem{Muraishi2009}
H.~Muraishi, S.~Miyake and S.~Yanagita, \emph{{A stochastic view of the
  propagation of galactic cosmic rays}},  in \emph{Proceedings of the 31st ICRC
  Lodz 2009}, 2009.
\newblock \href{http://dx.doi.org/10.1017/CBO9781107415324.004}{DOI}.

\bibitem{Effenberger2012a}
F.~Effenberger, H.~Fichtner, K.~Scherer and I.~B{\"{u}}sching,
  \emph{{Anisotropic diffusion of Galactic cosmic ray protons and their
  steady-state azimuthal distribution}},
  \href{http://dx.doi.org/10.1051/0004-6361/201220203}{\emph{Astronomy {\&}
  Astrophysics} {\bf 547} (Nov., 2012) }.

\bibitem{Miyake2015}
S.~Miyake, H.~Muraishi and S.~Yanagita, \emph{{A stochastic simulation of the
  propagation of Galactic cosmic rays reflecting the discreteness of cosmic ray
  sources Age and path length distribution}},
  \href{http://dx.doi.org/10.1051/0004-6361/201424442}{\emph{Astronomy {\&}
  Astrophysics} {\bf 573} (Jan., 2015) }.

\bibitem{Kopp2014}
A.~{Kopp}, I.~{B\"usching}, M.~{Potgieter} and R.~{Strauss}, \emph{A stochastic
  approach to galactic proton propagation: Influence of the spiral arm
  structure},
  \href{http://dx.doi.org/http://dx.doi.org/10.1016/j.newast.2014.01.006}{\emph{New
  Astronomy} {\bf 30} (Jul., 2014) }.

\bibitem{kampert2013}
K.-H. {Kampert} et~al., \emph{{CRPropa 2.0 - A public framework for propagating
  high energy nuclei, secondary gamma rays and neutrinos}},
  \href{http://dx.doi.org/10.1016/j.astropartphys.2012.12.001}{\emph{Astropart.~Phys.}
  {\bf 42} (Feb., 2013) }.

\bibitem{Batista2016}
R.~{Alves Batista}, A.~{Dundovic}, M.~{Erdmann}, K.-H. {Kampert}, D.~{Kuempel},
  G.~{M{\"u}ller} et~al., \emph{Crpropa 3---a public astrophysical simulation
  framework for propagating extraterrestrial ultra-high energy particles},
  \href{http://dx.doi.org/10.1088/1475-7516/2016/05/038}{\emph{J.~Cosm.~and~Astr.~Phys.}
  {\bf 5} (May, 2016) }.

\bibitem{Wiebel-Sooth1997}
B.~{Wiebel-Sooth}, P.~L. {Biermann} and H.~{Meyer}, \emph{Cosmic rays. vii.
  individual element spectra: prediction and data}, {\emph{Astron. \& Astroph.}
  {\bf 330} (Feb., 1998) }, [\href{http://arxiv.org/abs/astro-ph/9709253}{{\tt
  astro-ph/9709253}}].

\bibitem{allard2007}
D.~{Allard}, E.~{Parizot} and A.~V. {Olinto}, \emph{{On the transition from
  galactic to extragalactic cosmic-rays: Spectral and composition features from
  two opposite scenarios}},
  \href{http://dx.doi.org/10.1016/j.astropartphys.2006.09.006}{\emph{Astropart.~Phys.}
  {\bf 27} (Feb., 2007) }.

\bibitem{stanev1993}
T.~{Stanev}, P.~L. {Biermann} and T.~K. {Gaisser}, \emph{{Cosmic rays. IV. The
  spectrum and chemical composition above 10 GeV}}, {\emph{Astron. \& Astroph.}
  {\bf 274} (Jul., 1993) }.

\bibitem{bbr2009}
P.~L. {Biermann}, J.~K. {Becker}, A.~{Meli}, W.~{Rhode}, E.~S. {Seo} and
  T.~{Stanev}, \emph{{Cosmic Ray Electrons and Positrons from Supernova
  Explosions of Massive Stars}},
  \href{http://dx.doi.org/10.1103/PhysRevLett.103.061101}{\emph{Phys.~Rev.~Lett.}
  {\bf 103} (Aug., 2009) }.

\bibitem{hoerandel2004}
J.~R. {H{\"o}randel}, \emph{{Models of the knee in the energy spectrum of
  cosmic rays}},
  \href{http://dx.doi.org/10.1016/j.astropartphys.2004.01.004}{\emph{Astropart.~Phys.}
  {\bf 21} (June, 2004) }.

\bibitem{giacinti2014}
G.~{Giacinti}, M.~{Kachelrie{\ss}} and D.~V. {Semikoz}, \emph{{Explaining the
  spectra of cosmic ray groups above the knee by escape from the Galaxy}},
  \href{http://dx.doi.org/10.1103/PhysRevD.90.041302}{\emph{Phys.~Rev.~D} {\bf
  90} (Aug., 2014) }.

\bibitem{Thoudam2016}
S.~{Thoudam}, J.~P. {Rachen}, A.~{van Vliet}, A.~{Achterberg}, S.~{Buitink},
  H.~{Falcke} et~al., \emph{{Cosmic-ray energy spectrum and composition up to
  the ankle: the case for a second Galactic component}},
  \href{http://dx.doi.org/10.1051/0004-6361/201628894}{\emph{Astron. \&
  Astroph.} {\bf 595} (Oct., 2016) }.

\bibitem{Kopp2012}
A.~Kopp, I.~B{\"{u}}sching, R.~Strauss and M.~Potgieter, \emph{{A stochastic
  differential equation code for multidimensional Fokker–Planck type
  problems}}, \href{http://dx.doi.org/10.1016/j.cpc.2011.11.014}{\emph{Computer
  Physics Communications} {\bf 183} (Mar., 2012) }.

\bibitem{Giacalone-Jokipii-1999}
J.~{Giacalone} and J.~R. {Jokipii}, \emph{{The Transport of Cosmic Rays across
  a Turbulent Magnetic Field}},
  \href{http://dx.doi.org/10.1086/307452}{\emph{Astroph.~J.} {\bf 520} (July,
  1999) }.

\bibitem{Qin-Shalchi-2016}
G.~{Qin} and A.~{Shalchi}, \emph{{Numerical Test of Different Approximations
  Used in the Transport Theory of Energetic Particles}},
  \href{http://dx.doi.org/10.3847/0004-637X/823/1/23}{\emph{Astroph.~J.} {\bf
  823} (May, 2016) }.

\bibitem{Desmond2001}
D.~J. Higham., \emph{An algorithmic introduction to numerical simulation of
  stochastic differential equations},
  \href{http://dx.doi.org/10.1137/S0036144500378302}{\emph{SIAM Review} {\bf
  43} (Aug., 2001) }.

\bibitem{1969ArRMA..32...29M}
A.~Marris and S.~Passman, \emph{{Vector fields and flows on developable
  surfaces}}, \href{http://dx.doi.org/10.1007/BF00253256}{\emph{Archive for
  Rational Mechanics and Analysis} {\bf 32} (Jan., 1969) }.

\bibitem{Ruffolo_2008}
D.~{Ruffolo}, P.~{Chuychai}, P.~{Wongpan}, J.~{Minnie}, J.~W. {Bieber} and
  W.~H. {Matthaeus}, \emph{{Perpendicular Transport of Energetic Charged
  Particles in Nonaxisymmetric Two-Component Magnetic Turbulence}},
  \href{http://dx.doi.org/10.1086/591493}{\emph{Astroph.~J.} {\bf 686} (Oct.,
  2008) }.

\bibitem{JAN12}
R.~{Jansson} and G.~R. {Farrar}, \emph{{A New Model of the Galactic Magnetic
  Field}},
  \href{http://dx.doi.org/10.1088/0004-637X/757/1/14}{\emph{Astroph.~J.} {\bf
  757} (Sept., 2012) }.

\bibitem{EFF11}
F.~{Effenberger}, H.~{Fichtner}, K.~{Scherer} and I.~{B{\"u}sching},
  \emph{{Anisotropic diffusion of Galactic cosmic ray protons and their
  steady-state azimuthal distribution}},
  \href{http://dx.doi.org/10.1051/0004-6361/201220203}{\emph{Astron. \&
  Astroph.} {\bf 547} (Nov., 2012) }.

\bibitem{CAS90}
J.~R. Cash and A.~H. Karp, \emph{A variable order runge-kutta method for
  initial value problems with rapidly varying right-hand sides},
  \href{http://dx.doi.org/10.1145/79505.79507}{\emph{ACM Trans. Math. Softw.}
  {\bf 16} (Sept., 1990) }.

\bibitem{Anderson1952}
T.~W. Anderson and D.~A. Darling, \emph{{Asymptotic theory of certain "Goodness
  of fit" criteria based on stochastic processes}},
  \href{http://dx.doi.org/10.1007/s13398-014-0173-7.2}{\emph{The Annals of
  Mathematical Statistics} {\bf 23} (June, 1952) }.

\bibitem{Blasi2012I}
P.~{Blasi} and E.~{Amato}, \emph{{Diffusive propagation of cosmic rays from
  supernova remnants in the Galaxy. I: spectrum and chemical composition}},
  \href{http://dx.doi.org/10.1088/1475-7516/2012/01/010}{\emph{J.~Cosm.~and~Astr.~Phys.}
  {\bf 1} (Jan., 2012) }.

\bibitem{Case1996}
G.~{Case} and D.~{Bhattacharya}, \emph{{Revisiting the galactic supernova
  remnant distribution.}}, {\emph{Astron. \& Astroph. Suppl.} {\bf 120} (Dec.,
  1996) }.

\bibitem{Heesen2009}
V.~{Heesen}, M.~{Krause}, R.~{Beck} and R.-J. {Dettmar}, \emph{{The magnetic
  field structure in NGC 253 in presence of a galactic wind}},  in \emph{Cosmic
  Magnetic Fields: From Planets, to Stars and Galaxies} (K.~G. {Strassmeier},
  A.~G. {Kosovichev} and J.~E. {Beckman}, eds.), vol.~259 of \emph{IAU
  Symposium}, Apr., 2009.
\newblock \href{http://dx.doi.org/10.1017/S1743921309031184}{DOI}.

\bibitem{Ferriere2014}
K.~{Ferri{\`e}re} and P.~{Terral}, \emph{{Analytical models of X-shape magnetic
  fields in galactic halos}},
  \href{http://dx.doi.org/10.1051/0004-6361/201322966}{\emph{Astron. \&
  Astroph.} {\bf 561} (Jan., 2014) }.

\bibitem{Lopez2016}
C.~{L{\'o}pez-Cob{\'a}}, S.~F. {S{\'a}nchez}, A.~V. {Moiseev}, D.~V. {Oparin},
  T.~{Bitsakis}, I.~{Cruz-Gonz{\'a}lez} et~al., \emph{{Star formation driven
  galactic winds in UGC 10043}},
  \href{http://dx.doi.org/10.1093/mnras/stw3355}{\emph{Mon.~Not.~Roy.~Astron.~Soc.}
  (Dec., 2016) }.

\bibitem{FAR14}
G.~R. {Farrar}, \emph{{The Galactic magnetic field and ultrahigh-energy cosmic
  ray deflections}},
  \href{http://dx.doi.org/10.1016/j.crhy.2014.04.002}{\emph{Comptes Rendus
  Physique} {\bf 15} (Apr., 2014) }.

\bibitem{Faucher2006}
C.-A. {Faucher-Gigu{\`e}re} and V.~M. {Kaspi}, \emph{{Birth and Evolution of
  Isolated Radio Pulsars}},
  \href{http://dx.doi.org/10.1086/501516}{\emph{Astroph.~J.} {\bf 643} (May,
  2006) }.

\bibitem{Ackermann2011}
M.~{Ackermann} et~al., \emph{{Constraints on the Cosmic-ray Density Gradient
  Beyond the Solar Circle from Fermi {$\gamma$}-ray Observations of the Third
  Galactic Quadrant}},
  \href{http://dx.doi.org/10.1088/0004-637X/726/2/81}{\emph{Astroph.~J.} {\bf
  726} (Jan., 2011) }.

\bibitem{Mazziotta2016}
M.~N. {Mazziotta}, F.~{Cerutti}, A.~{Ferrari}, D.~{Gaggero}, F.~{Loparco} and
  P.~R. {Sala}, \emph{{Production of secondary particles and nuclei in cosmic
  rays collisions with the interstellar gas using the FLUKA code}},
  \href{http://dx.doi.org/10.1016/j.astropartphys.2016.04.005}{\emph{Astroparticle
  Physics} {\bf 81} (Aug., 2016) }.

\bibitem{Webber2003}
W.~R. {Webber}, A.~{Soutoul}, J.~C. {Kish} and J.~M. {Rockstroh},
  \emph{{Updated Formula for Calculating Partial Cross Sections for Nuclear
  Reactions of Nuclei with Z $\leq$ 28 and E $\geq$ 150 MeV Nucleon$^{-1}$ in
  Hydrogen Targets}},
  \href{http://dx.doi.org/10.1086/344051}{\emph{Astroph.~J.~Suppl.~Series} {\bf
  144} (Jan., 2003) }.

\bibitem{Nakanishi2016}
H.~{Nakanishi} and Y.~{Sofue}, \emph{{Three-dimensional distribution of the ISM
  in the Milky Way galaxy. III. The total neutral gas disk}},
  \href{http://dx.doi.org/10.1093/pasj/psv108}{\emph{Publications of the ASJ}
  {\bf 68} (Feb., 2016) }.

\bibitem{Snodin2016}
A.~P. Snodin, A.~Shukurov, G.~R. Sarson, P.~J. Bushby and L.~F.~S. Rodrigues,
  \emph{{Global diffusion of cosmic rays in random magnetic fields}},
  \href{http://dx.doi.org/10.1093/mnras/stw217}{\emph{Monthly Notices of the
  Royal Astronomical Society} {\bf 457} (Apr., 2016) }.

\bibitem{van_der_Walt_2011}
S.~{van der Walt}, C.~S. {Colbert} and G.~{Varoquaux}, \emph{The numpy array: A
  structure for efficient numerical computation},
  \href{http://dx.doi.org/10.1109/MCSE.2011.37}{\emph{Computing in Science \&
  Engineering} {\bf 13} (2011) }.

\bibitem{McKinney_2012}
W.~McKinney, \emph{Python for Data Analysis}.
\newblock O'Reilly, 1~ed., 2012.

\bibitem{Hunter_2007}
J.~D. Hunter, \emph{Matplotlib: A 2d graphics environment},
  \href{http://dx.doi.org/10.1109/MCSE.2007.55}{\emph{Computing In Science \&
  Engineering} {\bf 9} (2007) }.

\bibitem{Perez_2007}
F.~P{\`e}rez and G.~B. E, \emph{Ipython: A system for interactive scientific
  computing}, \href{http://dx.doi.org/10.1109/MCSE.2007.53}{\emph{Computing in
  Science \& Engineering} {\bf 9} (2007) }.

\end{thebibliography}\endgroup
